\documentclass[aps,prc,twocolumn,showpacs,superscriptaddress]{revtex4-2}
\usepackage{natbib}
\usepackage{graphicx}
\usepackage{color}
\usepackage{physics}
\usepackage{amsfonts}
\usepackage{braket}
\usepackage{subfig}

\usepackage{float}
\makeatletter
\let\newfloat\newfloat@ltx
\makeatother

\usepackage{algorithm}
\usepackage{algpseudocode}
\usepackage{amsmath, amssymb, mathrsfs}

\begin{document}

\title[Article Title]{Solution of the Equation-of-Motion Phonon Method eigenvalue problems on the D-Wave quantum annealer}
\author{C. De Lucia}
\affiliation{Centro Italiano Ricerche Aerospaziali (CIRA), Via Maiorise, 81043 Capua, CE, Italy;}
\author{A. Martone}
\affiliation{Centro Italiano Ricerche Aerospaziali (CIRA), Via Maiorise, 81043 Capua, CE, Italy;}
\author{F.A. D'Aniello}
\affiliation{Centro Italiano Ricerche Aerospaziali (CIRA), Via Maiorise, 81043 Capua, CE, Italy;}
\author{A. Mastroianni}
\affiliation{Dipartimento di Matematica e Fisica, Universit\`a degli
  Studi della Campania ``Luigi Vanvitelli'', viale Abramo Lincoln 5 -
  I-81100 Caserta, Italy}
\author{G. Nunziata}
\affiliation{Dipartimento di Matematica e Fisica, Universit\`a degli
  Studi della Campania ``Luigi Vanvitelli'', viale Abramo Lincoln 5 -
  I-81100 Caserta, Italy}
\author{G. De Gregorio}
\affiliation{Dipartimento di Matematica e Fisica, Universit\`a degli
  Studi della Campania ``Luigi Vanvitelli'', viale Abramo Lincoln 5 -
  I-81100 Caserta, Italy}
\affiliation{Istituto Nazionale di Fisica Nucleare, \\ 
Complesso Universitario di Monte  S. Angelo, Via Cintia - I-80126 Napoli, Italy}
\author{R. Folprecht}   
\affiliation{Institute of Particle and Nuclear Physics, Faculty of Mathematics and Physics,  Charles University, V Hole\v sovi\v ck\'ach 2, 180 00 Prague, Czech Republic } 
\author{F. Knapp}   
\affiliation{Institute of Particle and Nuclear Physics, Faculty of Mathematics and Physics,  Charles University, V Hole\v sovi\v ck\'ach 2, 180 00 Prague, Czech Republic } 
\author{N. Lo Iudice}
\affiliation{Dipartimento di Fisica, 
Universit\`a di Napoli Federico II, 80126 Napoli, Italy} 
\author{P. Vesel\'y} 
\affiliation{Nuclear Physics Institute,
Czech Academy of Sciences, 250 68 \v Re\v z, Czech Republic} 

\begin{abstract}

The solution of large-scale eigenvalue problems is crucial in nuclear many-body theory, where Hamiltonian matrices often reach extremely large dimensions. Quantum computing opens new perspectives for addressing such demanding problems. 
Although the Quantum Phase Estimation algorithm offers, in principle, a systematic route to matrix diagonalization, its practical deployment demands levels of coherence and error correction that current quantum hardware cannot yet support. A viable near-term strategy is instead to exploit quantum annealing, which enables the recasting of eigenvalue problems into quadratic unconstrained binary optimization formulations that can be addressed by existing annealing-based processors. Here, we propose a hybrid quantum-classical algorithm that combines quantum annealing and classical deflation to iteratively extract the full eigenspectrum of both standard and generalized eigenvalue problems. We benchmark this method on eigenvalue problems arising from the Equation of Motion Phonon Method performing calculations on real quantum hardware. Our approach illustrates the capabilities and limitations of near-term quantum devices in addressing nuclear eigenvalue problems.
\end{abstract}


\maketitle
\section{Introduction} 
\label{sec:introduction}
The nuclear many-body problem cannot be solved exactly for systems with more than a few nucleons \cite{FB0,FB1,FB2,FB4,FB5}, due to the extremely rapid growth of the Hilbert space dimension. As a consequence, a wide variety of many-body methods has been developed, each introducing controlled approximations to make the problem computationally tractable. The size of the Hamiltonian matrices resulting from these approaches depends both on the chosen truncation scheme and on the number of active nucleons; nevertheless, even with sophisticated approximations, the resulting matrices can reach extremely large dimensions \cite{Shimizu2022}. Among others we mention the \textit{ab initio} methods, such as Coupled-Cluster (CC) theory \cite{CC1,CC2} and the In-Medium Similarity Renormalization Group (IM-SRG) \cite{IMRSG1,IMRSG2,IMRSG3}, the random-phase approximation and its extensions \cite{SRPA1,SRPA2,SRPA3,SRPA4}, the Relativistic Time-Blocking Approximation (RTBA) \cite{RTBA1,RTBA2,RTBA3}, and, finally, the method adopted in this work, the Equation of Motion Phonon Method (EMPM)~\cite{AndLo1}. 

This approach has been successfully applied to study low- and high-energy spectroscopic properties of even-even~\cite{DeGregorio21,DeGregorio22} and odd-even nuclei~\cite{DeGreg16a,DeGreg17}, as well as extended to the quasiparticle basis for open-shell systems~\cite{DeGreg16}. The EMPM mitigates the challenges of large-scale many-body nuclear problems by constructing an orthonormal basis of multiphonon states built upon one-phonon excitations computed in the Tamm-Dancoff Approximation (TDA). By solving a set of equations of motion, EMPM captures collective correlations up to arbitrary phonon number while compressing the Hamiltonian into smaller blocks. Nevertheless, these matrices can still reach very large dimensions, and repeated diagonalizations remain computationally costly in large-scale calculations.

Classical iterative eigensolvers-such as the Lanczos \cite{Lanczos1,Lanczos2} and Davidson algorithm \cite{Davidson1975}-exploit sparsity and subspace projection to compute extremal eigenpairs without constructing the full spectrum~\cite{Saad11}. These methods build progressively larger Krylov or Davidson subspaces and converge rapidly when spectral gaps are large. However, the storage of growing subspaces becomes a bottleneck, even with modern high-performance implementations \cite{Saad11,Lanczos2}.

Quantum computing offers an alternative route to solve eigenvalue problems. In principle, the Quantum Phase Estimation (QPE) algorithm~\cite{Nielsen10} can extract eigenvalues of Hermitian operators with arbitrary precision, provided fault-tolerant qubits and deep circuits are available. Despite its theoretical potential, QPE remains out of reach for current Noisy Intermediate-Scale Quantum (NISQ) devices due to coherence and error correction limitations~\cite{Preskill2018}.

To bridge the gap between quantum algorithms and today's noisy hardware, many hybrid approaches have been developed. Most of these methods were initially designed to approximate the ground-state energy of a quantum system, with the Variational Quantum Eigensolver (VQE) and its numerous extensions playing a central role \cite{VQE1,VQE2}. Other hybrid formulations, such as the quantum version of the Lanczos algorithm \cite{QLancz}, were likewise introduced with the primary goal of improving ground-state calculations.
Building on the success of ground-state VQE, extensions aimed at computing excited states were subsequently proposed \cite{VQEext1,VQEext2,VQEext3}. Despite their practical relevance, these approaches inherit several limitations of the variational framework: the optimization landscape is highly nonlinear, the energy estimators are affected by stochastic noise, and the complexity grows significantly when multiple excited states are included.  Further hybrid strategies based on the Davidson algorithm have also been explored for excited-state targeting \cite{QDavidson}. 

An alternative and increasingly influential direction for excited-state calculations on quantum computers is based on quantum subspace methods. This class of approaches includes quantum subspace expansion \cite{EXC1,EXC2,EXC3}, non-orthogonal VQE \cite{NOVQE}, equation-of-motion techniques \cite{QEOM1,QEOM2,QEOM3,QEOM4}, and the Quantum Krylov Subspace (QKS) framework, inspired by classical Krylov algorithms \cite{QKrylov1,QKrylov2,QKrylov3}. These methods aim to overcome some of the limitations of standard variational algorithms by projecting the problem onto a suitably constructed low-dimensional subspace, often resulting in more stable and scalable excited-state computations.

Another hybrid approach is based on quantum annealing and Quadratic Unconstrained Binary Optimization (QUBO) problems. 
Recent works have demonstrated the feasibility of quantum annealing for addressing both Standard (SEVPs) and Generalized EigenValue Problems
(GEVPs)~\cite{Krakoff22,Illa2022,TostiBalducci2022,Stockley2023}. In particular, the authors in~\cite{Krakoff22} introduced an iterative steepest-descent approach where each step of Rayleigh quotient minimization is encoded into a QUBO instance, solved via quantum annealing. Repeated annealing steps drive convergence toward the ground-state eigenpair of a symmetric matrix, as shown in proof-of-principle demonstrations on small-scale problems~\cite{Rajak23}. 
However, this scheme is intrinsically limited to computing only the lowest-energy solution.

Building on this foundation, the authors in~\cite{Fornetti} demonstrated that it is possible to compute the minimal or maximal right eigenvalue-and corresponding eigenvector-of nonsymmetric GEVPs using a QUBO formulation compatible with quantum annealers. Their specific target stems from the discretization of the homogeneous Bethe-Salpeter equation (hBSE) in ladder approximation, formulated directly in four-dimensional Minkowski space. This development shows that quantum annealers, such as the D-Wave Advantage system with more than 5,000 qubits and 15-way connectivity, can be employed not only to solve symmetric GEVPs but also to tackle nonsymmetric problems relevant to relativistic bound-state equations.

In this work, we build upon the algorithm outlined in~\cite{Krakoff22} and extend it to recover the full eigenspectrum for both SEVPs and GEVPs. We present a hybrid classical-quantum framework where successive eigenpairs are extracted by alternating QUBO-based quantum optimizations with classical deflation steps. After obtaining the \(k\)-th eigenpair via quantum annealing, we project out the span of previously found eigenvectors, allowing the next QUBO call to target the subsequent eigenstate in descending order of energy magnitude.

We benchmark our extended algorithm on Hamiltonian matrices arising from EMPM calculations in nuclear structure. By varying parameters such as the harmonic-oscillator basis size and binary encoding precision, we compare simulated annealing and quantum annealing in terms of convergence behaviour, eigenvalue accuracy, and scalability. This work provides a practical strategy for exploiting near-term quantum hardware in nuclear-structure calculations. By combining classical preprocessing with quantum annealing strategies in a scalable framework, our approach opens new directions for solving large nuclear eigenvalue problems, moving closer to exploiting quantum computational resources for realistic physical applications.

\section{The EMPM: A brief outline}
Assuming that the $(n-1)$-phonon basis states $\ket{\alpha_{(n-1)}}$ of energies $E_{\alpha_{(n-1)}}$ are known,  we construct the set of redundant states 
 \begin{equation}
 \ket{\lambda \alpha_{n-1}}  =  O^\dagger_\lambda \ket{\alpha_{n-1}},
 \end{equation}
where
\begin{equation}
\label{lam}
 O^\dagger_\lambda  = \sum_{ph} c^\lambda_{ph} a^\dagger_p   b_h,
\end{equation}
creates a TD phonon of energy E$_\lambda$ out of HF vacuum $\ket{0}$ through the action of the particle
 ($a^\dagger_p=  a^\dagger_{x_p j_p m_p} $) and hole ($b_h = (-)^{j_h + m_h} a_{x_h j_h - m_h}$) creation operators.
  
 We first extract from the redundant set a basis of linearly independent (though not orthogonal) states  $\ket{\lambda \alpha_{n-1}} $ using the Cholesky decomposition method
 and use this basis to derive and solve the eigenvalue problem within the $n$-subspace. To this end we start from the equations of motion
 \begin{equation}
\bra{\alpha_{n-1}} [O_\lambda,H] \ket{\alpha_n}=  
(E_{\alpha_n} - E_{\alpha_{n-1}}) \langle  \lambda \alpha_{n-1} \ket{\alpha_n}.
\label{EoM}
\end{equation}
After expanding the commutator and carrying out the standard algebraic manipulations detailed in Ref. \cite{DeGregorio22}, we obtain the generalized eigenvalue equations
\begin{equation}
\Bigl( {\cal H } - E   \Bigr) D  C =0,
\end{equation}
or more explicitly 
\begin{equation}
\label{eq:EMPM}
\sum_{jk} \Bigl({\cal H} ^{\alpha_n}_{ik}  - E_{\alpha_n} \delta_{ik} \Bigr)
{\cal D}^{\alpha_n}_{kj}C^{\alpha_n}_{j}=  0.
\end{equation}
 Here
\begin{eqnarray}
{\cal H} ^{\alpha_n}_{ik}  ={\cal H} ^{\alpha_n}_{(\lambda \alpha_{n-1}) (\lambda" \alpha"_{n-1})}  = \nonumber\\
(E_\lambda  + E_{\alpha_{n-1}} ) \delta_{\lambda \lambda"} \delta_{\alpha_{n-1} \alpha"_{n-1}} 
+ {\cal V}^{\alpha_n}_{(\lambda \alpha_{n-1}) (\lambda" \alpha"_{n-1})},
\label{Hn}
 \end{eqnarray}
 where  ${\cal V}^{\alpha_n}_{(\lambda \alpha_{n-1}) (\lambda" \alpha"_{n-1})}$ 
defines the phonon-phonon interaction, and
\begin{equation}
\label{D}
{\cal D}^{\alpha_n}_{ij} =  {\cal D}^{\alpha_n}_{(\lambda \alpha_{n-1}) (\lambda' \alpha'_{n-1})}= \braket{\lambda ' \alpha'_{n-1} \mid \lambda \alpha_{n-1}},
\end{equation}
is the overlap or metric matrix which preserves   the Pauli principle. The expressions of ${\cal D}$ and ${\cal V}$  can be found, for instance, in Ref. \cite{DeGregorio22}.

At this stage, we employ the singular value decomposition (SVD)  to identify and remove spurious states \cite{DeGregorio21}. The intrinsic states obtained after this projection satisfy the transformed eigenvalue equation 
\begin{equation}
\Bigl( {\cal H'} - E \Bigr) D' C'=0.
\label{eign}
\end{equation}
The c.m. free $n$-phonon eigenstates so obtained  can be recast in terms of the original basis
\begin{equation}
\ket{\alpha_n}  = \sum_{\lambda \alpha_{n-1}}  C_{\lambda \alpha_{n-1} }^{\alpha_n} \ket{\lambda \alpha_{n-1}}.
\label{nstate}
\end{equation}
By iterating this procedure for successive values of $n$, we generate a complete orthonormal set of multiphonon state which,  added to  the HF vacuum ($\ket{0}$) and  the TD phonons ($\{\ket{ \alpha_1}\} = \{\ket{ \lambda}\}$), form an orthonormal basis $\{\ket{\alpha_n}\}$ ($n=0,1,2,3,...$).

This basis is then used to construct and solve the eigenvalue problem in the full multiphonon space
\begin{equation}
\label{eq:eigfull}
\sum_{ \alpha_n \beta_{n'}} \Bigl((E_{\alpha_n} - {\cal E}_\nu) \delta_{\alpha_n \beta_{n'}}  +   {\cal V}_{\alpha_n \beta_{n'}} \Bigr){\cal C}^{\nu}_{\beta_{n'}} = 0,
\end{equation}
where ${\cal V}_{\alpha_n \beta_{n'}}  = 0$ for  $n' = n$.

The coupling of the $n$-phonon $\ket{\alpha} =\ket{\alpha_n} $ to the $n'$-phonon  states $\ket{\beta'} = \ket{\beta_{n'}}$    
has the structure  
\begin{equation}
\label{couponen}
{\cal V}_{\alpha \beta'} =     
   \sum_{ \lambda' \alpha'}    {\cal V}^{\lambda'}_{\alpha \alpha'}   \langle \lambda' \alpha' \ket{ \beta' },
\end{equation}
for $n'=n+1$ and 
\begin{equation}
{\cal V}_{\alpha \beta'} = \sum_{\alpha_2} \bra{0}  H \ket{\alpha_2 } \langle \alpha_2 \alpha  \ket{ \beta'}
\label{couptwon}
\end{equation}
for $n'=n+2$.
The formulas giving ${\cal V}^{\lambda'}_{\alpha \alpha'} $ and $ \bra{0}  H \ket{\alpha_2 }$,
can be found elsewhere \cite{DeGregorio22}.
The solution of the final eigenvalue Eq. \eqref{eq:eigfull} yields the eigenvectors ($n=0,1,2,3....$)
\begin{equation}
\label{Psifull}
\ket{ \Psi_\nu } = \sum_{n,\alpha_n}  {\cal C}_{\alpha_n}^\nu \ket{ \alpha_n }.
\end{equation}

\section{Qubo-based algorithm}\label{sec:Qubo_based_algorithm}
Having outlined the EMPM formalism, we now describe the QUBO-based algorithm used to solve the associated eigenvalue problems. Let $A$ be a symmetric matrix, a well-known consequence of the spectral theorem is that the smallest eigenvalue $\lambda$ and the corresponding eigenvector $v$ are global minima for the Rayleigh quotient $\bra{x}A\ket{x}/\braket{x|x}$, therefore 

\begin{equation}
\lambda=\min _{\|x\|=1} \bra{x} A \ket{x}, \quad \ket{v}=\underset{\|x\|=1}{\text{argmin}} \bra{x} A \ket{x}.
\end{equation}
Equivalently, one can search for the maximum magnitude of the Rayleigh quotient. Thus, for given a real symmetric matrix \( A \), one has to compute
\begin{equation}
\lambda=\max _{\|x\|=1} \lvert\bra{x} A \ket{x}\rvert, \quad \ket{v}=\underset{\|x\|=1}{\arg\max} \lvert\bra{x} A \ket{x}\rvert.
\end{equation}

The maximum magnitude of a function can be found by evaluating both its minimum and maximum values and taking the one with the largest absolute value. Moreover, the maximum of a function can be conveniently computed by minimizing its negation.

To encode this minimization problem on a quantum annealer, the continuous Rayleigh quotient optimization must be reformulated in terms of binary variables to both obtain a good initial guess for the global minimum, and to implement an iterative descent from this initial estimate.

To address a real-variable optimization problem, one can encode it in the QUBO form by first approximating each real variable $x$ with  $b$ binary variables. Once the number of bits $b$ has been set, we can build the precision vector of length $b$

\begin{equation}
    \bra{p} = \left( -1, \frac{1}{2}, \frac{1}{2^2}, \frac{1}{2^3}, \ldots, \frac{1}{2^{b-1}} \right),
\end{equation}
the real variable $x$ can be approximated as
\begin{equation}
    x = \braket{ p| x_b},
\end{equation}
where $\ket{x_b}$ is a binary vector of dimension $b$. So, the set of all the real values we can obtain with a precision vector of dimension $b$ is 
\begin{equation}
C_{1, b}:=\left\{ \braket{p | x_b} \mid \ket{x_b} \in\{0,1\}^{b}\right\}.
\end{equation}

More generally, the $n$-fold product of $C_{1,b}$ is the set

\begin{equation}
C_{n, b}:=\left\{\left(I_n \otimes \ket{p}\right) \ket{x_b} \mid \ket{x_b} \in\{0,1\}^{n b}\right\},
\end{equation}

where $I_n$ is the identity matrix and the set $C_{n, b}$ is a discretized cube. 
Now, let $Q$ be a symmetric $n \times n$ matrix, $\ket{r}$ an $n$-vector and suppose we want to solve the constrained quadratic programming problem

\begin{equation}\label{eq: constrained quadratic programmin problem}
\underset{\ket{x}\in[-1,1]^n}{\text{argmin }} \braket{r \mid x}+ \bra{x} Q \ket{x}.
\end{equation}

To get an approximate solution to Eq. \eqref{eq: constrained quadratic programmin problem} using a QUBO, we first replace the unit cube by a discretized unit cube to get the optimization problem

\begin{equation}\label{eq: setting the qubo problem}
\underset{x \in C_{n,b}}{\text{argmin }} \braket{ r \mid x} + \bra{x} Q \ket{x}.
\end{equation}

Setting $m = nb$, $\ket{x} = (I_n \otimes \ket{p})\ket{x_b}$, and defining the precision matrix as 

\begin{equation}
    P = \ket{{p}} \bra{p} = \left(\begin{array}{lll}
\bra{p} & & \\
& & \\
& \bra{p} & \\
& & \\
& & \bra{p}
\end{array}\right),
\end{equation}

the Eq. \eqref{eq: setting the qubo problem} is equivalent to the following QUBO problem

\begin{equation}\label{eq: qubo for the unconstrained programming problem}
\begin{aligned}
& \underset{x_b \in\{0,1\}^m}{\operatorname{argmin}} \bra{r}\left(I_n \otimes \ket{p}\right) \ket{x_b} + \bra{x_b} (Q \otimes P) \ket{x_b} = \\
& =\underset{x_b \in\{0,1\}^m}{\operatorname{argmin}} \bra{x_b} \operatorname{Diag}\left(\bra{3} I_n \otimes \ket{p}\right) \ket{x_b} + \bra{x_b} (Q \otimes P) \ket{x_b} \\
& =\underset{x_b \in\{0,1\}^m}{\operatorname{argmin}} \bra{x_b} \left(\operatorname{Diag}\left(\bra{r} I_n \otimes \ket{p}\right)+ Q \otimes P\right) \ket{x_b}
\end{aligned}
\end{equation}

where Diag$(\ket{v})$ refers to the diagonal matrix with entries from $\ket{v}$.
Therefore, given a number of bits $b$, this procedure approximates the real optimization problem in $n$ variables of Eq. \eqref{eq: constrained quadratic programmin problem} with the QUBO of Eq. \eqref{eq: qubo for the unconstrained programming problem} of size $n \cdot b$. We conclude by highlighting that we are not restricted to the cube $\left[ -1, 1 \right]^n$, but if we want to optimize over the cube $\left[ -\delta, \delta \right]^n$, we need to repeat the same construction with the precision vector $\delta \cdot \ket{p}$.

Algorithms that solve continuous optimization problems rely on a good initial guess and an iterative descent rule. These tasks can be formulated as QUBO problems when trying to minimize the Rayleigh quotient over the unit sphere.

\subsection{Initial guess phase}
As in most iterative optimization procedures, a suitable initial guess is required to ensure convergence of the descent phase. 
Suppose that $\lambda_1 \le \lambda_2 \le \ldots \le \lambda_n $ are the eigenvalues of $A$. To get an initial approximation of $\lambda_1$, one can ask to solve

\begin{equation}\label{eq: qubo problem to solve}
\underset{x \in C_{n,b}}{\text{argmin }} 
 \bra{x} A \ket{x},
\end{equation}

as approximation of 

\begin{equation}
\underset{||x||=1}{\text{argmin }} 
 \bra{x} A \ket{x}.
\end{equation}

To avoid the collapse of the solution to $\ket{x} = \ket{0}$ when $A$ is positive definite we replace $A$ with $A - \lambda I_n$, where $\lambda \in (\lambda_i,\lambda_{i+1})$ for some $i$. This spectral shift preserves the eigenvectors, while the eigenvalues can be recovered from the new matrix

\begin{equation}\label{eq: qubo to solve 2}
\underset{\ket{x} \in C_{n,b}}{\text{argmin }} 
 \bra{x} \left( A-\lambda I_n  \right) \ket{x}.
\end{equation}

A relatively good initial choice for $\lambda$ is the average of the eigenvalues $\lambda = \text{tr}(A) / n $ as suggested in \cite{Krakoff22}. This is a natural choice since it ensures that the initial guess lies within the bounds of the eigenspectrum. Nevertheless, this estimate can be refined. In fact, following the statistical analogy introduced in \cite{Wolkowicz1980}, one defines the mean $m$ and standard deviation $s$ of the spectrum as

\begin{equation}
m=\frac{Tr(A)}{n}=\frac{1}{n}\sum_{j=1}^n \lambda_j,
\end{equation}

\begin{equation}
s^2=\frac{1}{n}Tr(A^2)-m^2.
\end{equation}
As a consequence, 

\begin{eqnarray}
 m-s\sqrt{n-1} \le \lambda_{min} \le m-\frac{s}{\sqrt{n-1}},\nonumber \\ \quad m+\frac{s}{\sqrt{n-1}} \le \lambda_{max} \le m+s\sqrt{n-1}, 
\end{eqnarray}

that for n=2 reduces to 
 \begin{eqnarray}
     \lambda_{min} = m-s,  \qquad\lambda_{max} = m + s.
 \end{eqnarray}

In general we have that,
\begin{equation}
m-s\sqrt{\frac{k-1}{n-k+1}} \le \lambda_{k} \le m+s\sqrt{\frac{n-k}{k}}.
\end{equation}

Once set the initial value for the eigenvalue, as the algorithm progresses, it decreases towards the target $\lambda_1$. 

These observations lead to the following algorithm, based on a iterative fixed-point method.

The algorithm starts with the initial guess phase, where we first solve Eq. \eqref{eq: qubo to solve 2} to produce a guess eigenvector $\ket{v_1}$. Then, we update $\lambda$ using the Rayleigh quotient $\lambda = \frac{\bra{v_1} A \ket{v_1}}{||v_1||^2}$ and we solve Eq. \eqref{eq: qubo problem to solve} again using $\ket{v_1}$ as a bias vector to produce a second guess vector $\ket{v_2}$. This procedure is repeated until $\lambda$ is no longer decreasing. 

\subsection{Iterative descent phase}
After obtaining an initial approximation of the lowest eigenvalue, the algorithm refines it through an iterative descent procedure.\\

Suppose we want to minimize a function $f: \mathbb{R}^m \rightarrow \mathbb{R}$, and let $\nabla f$ and $\mathcal{H}e(f)$ denote the gradient and Hessian of $f$, respectively. Starting with an initial guess $\ket{x_0}$, a common strategy is to Taylor expand $f$ around $\ket{x_0}$, namely

\begin{equation}
f(\ket{x})=f\left(\ket{x_0}\right)+\nabla f \cdot \ket{\delta} +\bra{\delta} \frac{\mathcal{H}e(f)}{2} \ket{\delta} + o\left(\|\delta\|^3\right),
\end{equation}

where $\ket{\delta} := \ket{x} - \ket{x_0}$, and choose a $\ket{\delta}$ to minimize $\nabla f \cdot \ket{\delta}$ (gradient descent), or $\nabla f \cdot \ket{\delta} + \bra{\delta} \frac{\mathcal{H}e(f)}{2} \ket{\delta}$ (second order methods such as Newton's method, BFGS, and Newton-CG \cite{NumOpt}). Once a better solution $\ket{x_1} = \ket{x_0} + \ket{\delta}$ has been found, we Taylor expand around $\ket{x_1}$ again and repeat. The proposed algorithm obtains a good descent direction by using a QUBO to find good approximate solutions to

\begin{equation}
\underset{\delta \in C_{n, b}}{\operatorname{argmin}} \nabla f \cdot \ket{\delta} + \bra{\delta} \frac{\mathcal{H}e(f)}{2} \ket{\delta}.
\end{equation}

Similar to Newton-CG and BFGS, this method requires computing, but not inverting, the Hessian matrix $\mathcal{H}e(f)$, and benefits from a line search which possibly increases the size of $\delta$.
If the $k$-th approximate solution $\ket{x_k}$ is closer to the true solution than any point in the discretized cube, one cannot expect minimizing the QUBO to produce a better solution. If the new candidate solution $\ket{x_{k+1}}$ is worse than the previous one, the algorithm discards the candidate and replaces the discretized unit cube $C_{n,b}$ by a scaled-down
discretized cube $t \cdot C_{n,b}$, where $t \ll 1$, which amounts to repeating the procedure outlined above with the new precision vector $t \ket{p}$. Once the discretized cube has been scaled down, the algorithm continues running until it needs to scale down the cube further, or terminates once the desired accuracy is reached.

\subsection{Extension to the generalized eigenvalue problem}\label{sec: adaptation to generalized eigenvalue problem}
In many cases, one needs to solve a generalized eigenvalue problem of the form $A \ket{v} = \lambda B \ket{v}$, where $A$, $B$ are symmetric and $B$ is Strictly Positive Definite (SPD). Under these conditions, the smallest generalized eigenvalue minimizes the generalized Rayleigh quotient

\begin{equation}
\lambda=\min _{\|x\|=1} \frac{\bra{x} A \ket{x}}{\bra{x} B \ket{x}}, \quad \ket{v}=\underset{\|x\|=1}{\operatorname{argmin}} \frac{\bra{x} A \ket{x}}{\bra{x} B \ket{x}}.
\end{equation}

Therefore, the following small changes adapt the algorithm discussed so far to solve the generalized eigenvalue problem with essentially arbitrary precision. First, instead of $m = \text{tr}(A) / n $, one can generate a random unit vector $\ket{w}$ and  compute $m$ as $m = \frac{\bra{w} A \ket{w}}{\bra{w} B \ket{w}}$. Second, we have to replace every Rayleigh quotient with the corresponding generalized Rayleigh quotient. Lastly, instead of updating $H$ as $H = A - \lambda I_n$, we update $H = A - \lambda B$, since this preserves the $B$-eigenspectrum of $A$, whereas the former does not.

\section{Matrix Deflation Techniques}\label{sec:Matrix_Deflation_Techniques}

The previously discussed algorithm computes a single eigenpair. To extend this approach to the computation of the full eigenspectrum, we employ matrix deflation techniques, which enable the sequential extraction of eigenvalues and eigenvectors.

Deflation is a classical strategy in numerical linear algebra: once an eigenpair has been computed, the system matrix is modified so that the same eigenpair does not reappear in subsequent iterations. This enables iterative procedures-such as the QUBO-based eigenvalue search described earlier-to identify additional eigenpairs one at a time.

Our use of deflation is motivated by the structure of the QUBO formulation. Unlike global eigensolvers (e.g., the QZ \cite{QZ0,QZ1} algorithm or dense QR methods \cite{QR1,QR2}), which compute multiple eigenpairs simultaneously, the QUBO solver isolates a single eigenpair per optimization instance. After each eigenpair is found, deflation ''peels away'' its influence from the matrix, allowing the algorithm to target the next dominant mode without interference.

Although a fully quantum deflation framework could, in principle, be implemented as part of a future quantum eigensolver, such an extension lies beyond the scope of this work.

\medskip

Deflation techniques can be broadly grouped into two main categories:
\begin{itemize}
    \item \textbf{Subtractive deflation:} The contribution of the computed eigenpair is subtracted from the matrix to eliminate its influence. Examples include Hotelling's deflation and Orthogonal Projection deflation.
    
    \item \textbf{Orthogonalization-based deflation:} The search space for subsequent eigenvector estimates is restricted to vectors orthogonal to the previously computed eigenvectors. Householder deflation is a representative method in this class.
\end{itemize}
\begin{table*}[t]
\centering
\begin{tabular}{|p{3.2cm}|p{4.6cm}|p{4.6cm}|p{5cm}|}
\hline
\textbf{Feature} & \textbf{Hotelling} & \textbf{Orthogonal Projection} & \textbf{Householder} \\
\hline
Basic Idea & Subtracts the rank-1 update formed by the found eigenpair from the matrix & Projects the matrix onto the subspace orthogonal to found eigenvectors & Uses Householder reflections to transform the matrix, removing found eigenvectors \\
\hline
Type & Subtractive & Subtractive & Orthogonalization-based \\
\hline
Mathematical Operation for SEVP& \( A' = A - \lambda \ket{v_1} \bra{v_1} \), where \(\lambda, \ket{v_1}\) are eigenvalue/vector & \( A' = (I - \ket{v_1} \bra{v_1}) ~A~ (I -  \ket{v_1} \bra{v_1}) \), where \(\ket{v_1}\) is the eigenvector & Applies orthogonal Householder transformation \( H \) to deflate, where \(H = I - 2 \frac{\ket{x} \bra{x}}{\|x\|_2^2}\), \(\ket{x} = \ket{v} \pm \|v\|_2 \ket{e_1}\) and $\ket{v}$ is the eigenvector\\
\hline
Mathematical Operation for GEVP& \( A' = A - \lambda_1 B \ket{v_1} \bra{v_1} B^T \), where \(\lambda, \ket{v_1}\) are eigenvalue/vector & \( A' = (I - B\ket{v_1} \bra{v_1})~ A~ (I - B\ket{v_1} \bra{v_1}) \), \( B' = (I - A\ket{v_1} \bra{v_1}) ~B~ (I - A\ket{v_1} \bra{v_1}) \), where $\ket{v_1}$ is the eigenvector & Applies orthogonal Householder transformation \( H \) to both  \(A \) and \(B \) matrices as follows:

\(A \;\leftarrow\; H\,A\,H,
  \qquad
  B \;\leftarrow\; H\,B\,H\)\\
\hline
Preserves Symmetry? & Yes & Yes & Yes \\
\hline
Numerical Stability & Moderate; can introduce errors due to subtraction & Higher stability due to projection & High stability from orthogonal transformations \\
\hline
Ease of Implementation & Simple & More complex; requires projection operations & More complex; requires constructing Householder reflectors \\
\hline
\end{tabular}
\caption{Comparison of Hotelling, Orthogonal Projection, and Householder deflation methods.}
\label{tab:deflation_comparison}
\end{table*}
In our procedure, we implement three classical and widely-used deflation techniques-Hotelling, Orthogonal Projection, and Householder deflation-covering both categories above. These methods are employed for both the standard and the generalized eigenvalue problems, enabling the sequential reconstruction of the full eigenspectrum as detailed in Section~\ref{sec:hybrid_algorithm_full_spectrum}. A summary of their main properties is provided in Table~\ref{tab:deflation_comparison}.

\subsection{Hotelling Deflation}\label{sec:Hotelling_Deflation}
Hotelling deflation is a technique used in eigenvalue problems to find additional eigenvalues after the dominant eigenpair (eigenvalue \(\lambda_1\) and eigenvector $\ket{v_1}$) has been found. It modifies the original matrix to "remove" the influence of the found eigenpair.
Given a matrix \( A \) and its dominant eigenpair \((\lambda_1, \ket{v_1})\), with $\ket{v_1}$ normalized so that \(\|v_1\| = 1\), the deflated matrix \( A' \) is defined as
\begin{equation}
A' = A - \lambda_1 \ket{v_1} \bra{v_1}.
\end{equation}

As result, the deflated matrix \( A' \) will have $\ket{v_1}$ as an eigenvector corresponding to an eigenvalue of zero. All other eigenvalues of \( A \) (and their corresponding eigenvectors) remain unchanged in \( A' \).
It is conceptually straightforward, but stability is its major weakness. Hotelling deflation is highly sensitive to errors in the computed \( \lambda_1 \) and $\ket{v_1}$. Small inaccuracies can accumulate rapidly, leading to poor accuracy for subsequently found eigenpairs. This makes it less suitable for accurately finding many eigenvalues.

To extend this deflation technique to the GEVP, the matrix \( B \) must be introduced. A common form of Hotelling-like deflation often involves modifying \( A \) while keeping \( B \) the same. If we have \( A \ket{v_1} = \lambda_1 B \ket{v_1} \), with$\ket{v_1}$ normalized such that \( \bra{v_1} B \ket{v_1} = 1 \), a Hotelling-like deflated matrix \( A' \) can be defined as
\begin{equation}
A' = A - \lambda_1 B \ket{v_1} \bra{v_1} B^T.
\end{equation}
Numerical instability, as in the standard problem, remains the primary issue, because the subtraction of terms like \( \lambda_1 B \ket{v_1} \bra{v_1} B^T\)  assumes perfect knowledge of \( \lambda_1 \) and \( \ket{v_1} \). In practice, computed eigenpairs are only approximate. Errors in \( \lambda_1 \) and \( \ket{v_1} \) are introduced at each deflation step, and these errors can accumulate rapidly.

\subsection{Orthogonal Projection Deflation}\label{sec:Orthogonal_Projection}
Orthogonal Projection deflation is a technique used in SEVP to find multiple eigenpairs by ensuring that subsequent eigenvector estimates are orthogonal to previously found eigenvectors. This is done by projecting the matrix onto the subspace orthogonal to the space spanned by the known eigenvectors.

Given a matrix \( A \) and its dominant eigenpair \((\lambda_1,\ket{v_1})\), with \(\ket{v_1} \) normalized so that \(\|v_1\| = 1\), the deflated matrix \( A' \) is defined as
\begin{equation}
A' = (I - \ket{v_1} \bra{v_1}) ~A~ (I -  \ket{v_1} \bra{v_1}),
\end{equation}
where \( I \) is the identity matrix.

This projection removes the influence of the known eigenvectors, allowing iterative methods to converge to new eigenpairs orthogonal to those already found. While conceptually useful and offering some advantages over the most basic Hotelling deflation, it is generally less numerically stable than Householder deflation and is not the primary method employed by state-of-the-art numerical libraries for computing complete eigenspectra (dense matrices) or subsets (sparse matrices).

For GEVP, Orthogonal Projection deflation removes the influence of already computed eigenvectors by projecting onto the subspace orthogonal to them with respect to the \( B \)-inner product. This is achieved by defining the projection operators
\begin{equation}
P_b = I - B  \ket{v_1} \bra{v_1}, \qquad P_a = I - A  \ket{v_1} \bra{v_1},
\end{equation}
and applying them to the matrices  \( A \) and  \( B \)
\begin{equation}
A' = P_b A P_b, \qquad B' = P_a B P_a.
\end{equation}

\subsection{Householder Deflation}\label{sec:Householder_Projection}
Householder deflation is another technique used in the SEVP to find additional eigenvalues, but employing Householder transformations. These transformations are orthogonal reflections that zero out components of vectors, allowing the matrix to be iteratively reduced in a numerically stable manner while preserving symmetry and orthogonality. By applying Householder deflation, the influence of already computed eigenvectors is effectively removed, which facilitates the computation of subsequent eigenpairs.
Given a matrix \( A \) and an eigenvector \( v_1 \), a Householder reflector \( H \) is constructed such that
\begin{equation}
H  \ket{v_1}  = \alpha  \ket{e_1},
\end{equation}
where \(  \ket{e_1} \) is the first canonical basis vector and \( \alpha  = \pm \| v_1 \|_2\). Applying \( H \) to \( A \) produces a transformed matrix
\begin{equation}
G = H A H^T.
\end{equation}
The transformed matrix \( G \) will have a block structure where \( \lambda_1 \) is isolated, and the remaining \((n-1) \times (n-1)\) submatrix \( A' \) contains the other eigenvalues of \( A \). Iterative deflation steps are then applied to \( A' \).

Householder deflation is numerically stable due to the orthogonality of \( H \) and is widely used in QR algorithms and advanced eigensolvers for SEVP.

For GEVP we apply Householder reflectors to both \( A \) and \( B \) matrices as follows
\begin{equation}\label{HouseholderTransofrmations}
  A \;\leftarrow\; H\,A\,H,
\qquad
  B \;\leftarrow\; H\,B\,H.
\end{equation}
However, as can be seen later in Section \ref{sec:FullSpectrum}, this approach distorts the original symmetry. Applying a single Householder transformation simultaneously to both \(A\) and \(B\) does not directly achieve deflation of an eigenpair in a clean and isolated manner. More specifically, such a transformation does not preserve the essential structural properties of the GEVP, such as maintaining \(B\) as a symmetric positive definite metric, or reducing \(A\) and \(B\) to triangular form. The objective of deflation is not only to eliminate the contribution of a specific eigenpair, but to do so while preserving the generalized eigenvalue structure and the mathematical relationship between \(A\) and \(B\). For this reason, modern approaches such as the generalized Schur decomposition (QZ algorithm) are preferred for the systematic computation of the eigenspectrum in the generalized case.

\section{Hybrid Algorithm for Full Spectrum Eigenpair Computation}\label{sec:hybrid_algorithm_full_spectrum}
In this section, we present a hybrid quantum-classical eigensolver algorithm designed to compute the full spectrum of eigenvalues and corresponding eigenvectors for both SEVP and GEVP. The proposed approach combines the algorithm presented in Section \ref{sec:Qubo_based_algorithm} with iterative eigenpair extraction via matrix deflation, using one of the deflation techniques described in Section \ref{sec:Matrix_Deflation_Techniques}. The algorithm is termed \textit{hybrid} as it combines a classical numerical procedure, matrix deflation, with a quantum-based formulation, namely the QUBO formulation of the Rayleigh quotient optimization problem.

\begin{algorithm}
\caption{Dominant eigenpair finder for the SEVP}
\label{max_magnitude_algorithm}
\begin{algorithmic}[1]
\Require Symmetric matrix \(A\)
\Ensure Eigenvalue \(\lambda_{\text{mm}}\) with maximum magnitude and corresponding eigenvector \(\ket{v_{\text{mm}}}\)
\State Solve QUBO to \textbf{maximize} the Rayleigh quotient:
\[
\ket{v_{\max}} = \underset{\|x\|=1}{\arg\max} \bra{x} A \ket{x}, \quad
\lambda_{\max} = \frac{\bra{v_{\max}} A \ket{v_{\max}}}{\braket{v_{\max} | v_{\max}}}
\]
\State Solve QUBO to \textbf{minimize} the Rayleigh quotient:
\[
\ket{v_{\min}} = \underset{\|x\|=1}{\arg\min} \bra{x} A \ket{x}, \quad
\lambda_{\min} = \frac{\bra{v_{\min}} A \ket{v_{\min}}}{\braket{v_{\min} | v_{\min}}}
\]
\If{\(|\lambda_{\max}| \geq |\lambda_{\min}|\)}
    \State \(\lambda_{\text{mm}} \gets \lambda_{\max}\), \(\ket{v_{\text{mm}}} \gets \ket{v_{\max}}\)
\Else
    \State \(\lambda_{\text{mm}} \gets \lambda_{\min}\), \(\ket{v_{\text{mm}}} \gets \ket{v_{\min}}\)
\EndIf
\State \Return \(\lambda_{\text{mm}}, \ket{v_{\text{mm}}}\)
\end{algorithmic}
\end{algorithm}

\begin{algorithm}
\caption{Hybrid full spectrum solver for the SEVP}\label{full_spectrum_eigensolver_sevp}
\begin{algorithmic}[1]
\Require Symmetric matrix $A$, and a chosen deflation method $\in \{$Hotelling, Orthogonal Projection, Householder$\}$
\Ensure Eigenvalues $\{\lambda_i\}$ and eigenvectors $\{\ket{v_i}\}$

\State Initialize set of computed eigenpairs: $\mathcal{S} \gets \emptyset$
\State Initialize deflated matrices: $(A_0) \gets (A)$

\For{$i = 1$ to desired number of eigenpairs}
   \State Find eigenpair (\(\lambda_i, \ket{v_i}\)) using Algorithm~\ref{max_magnitude_algorithm}:
    \[
    \lambda_i=\max _{\|x\|=1} \lvert\bra{x} A \ket{x}\rvert, \quad       \ket{v_i}=\underset{\|x\|=1}{\arg\max} \lvert\bra{x} A \ket{x}\rvert
    \]
    \State Add eigenpair to the set: $\mathcal{S} \gets \mathcal{S} \cup \{(\lambda_i, \ket{v_i})\}$
    \State Apply deflation to $(A_{i-1})$ to obtain $(A_i)$
\EndFor
\State \Return $\mathcal{S}$

\end{algorithmic}
\end{algorithm}

 \begin{algorithm}
\caption{Hybrid full spectrum solver for the GEVP}\label{full_spectrum_eigensolver_gevp}
\begin{algorithmic}[1]
\Require Symmetric matrix pair $(A, B)$, where $B$ is SPD, and a chosen deflation method $\in \{$Hotelling, Orthogonal Projection$\}$
\Ensure Eigenvalues $\{\lambda_i\}$ and eigenvectors $\{\ket{v_i}\}$

\State Initialize set of computed eigenpairs: $\mathcal{S} \gets \emptyset$
\State Initialize deflated matrices: $(A_0, B_0) \gets (A, B)$

\For{$i = 1$ to desired number of eigenpairs}
    \State Find eigenpair (\(\lambda_i, \ket{v_i}\)) using Algorithm~\ref{max_magnitude_algorithm}, extended for the GEVP:
    \[
    \lambda_i = \max _{\|x\|=1} \lvert \frac{\bra{x} A \ket{x}}{\bra{x} D \ket{x}}\rvert, \quad       \ket{v_i} = \underset{\|x\|=1}{\arg\max} \lvert\frac{\bra{x} A \ket{x}}{\bra{x} D \ket{x}}\rvert
    \]
    \State Add eigenpair to the set: $\mathcal{S} \gets \mathcal{S} \cup \{(\lambda_i, \ket{v_i})\}$
    \State Apply deflation to $(A_{i-1}, B_{i-1})$ to obtain $(A_i, B_i)$
\EndFor
\State \Return $\mathcal{S}$
\end{algorithmic}
\end{algorithm}

While in orthogonalization-based deflation (like Householder deflation), finding the minimum of the Rayleigh quotient at each iteration is sufficient to proceed, since the corresponding eigenpair is fully removed from the problem, this is not the case for subtractive deflation methods. In the latter, each iteration annihilates (but does not fully eliminate) the influence of the computed eigenpair. As a result, the same eigenpair might reappear as the minimum in subsequent iterations. If the matrix \(A\) is positive definite, this issue typically manifests already at the first deflation step. To overcome this issue, we employed the QUBO-based Algorithm~\ref{max_magnitude_algorithm}, to search for the maximum magnitude of the Rayleigh quotient, which corresponds to identifying the dominant eigenpair. 
\begin{figure*}[t]
\centering

\subfloat{
    \includegraphics[width=0.32\textwidth]{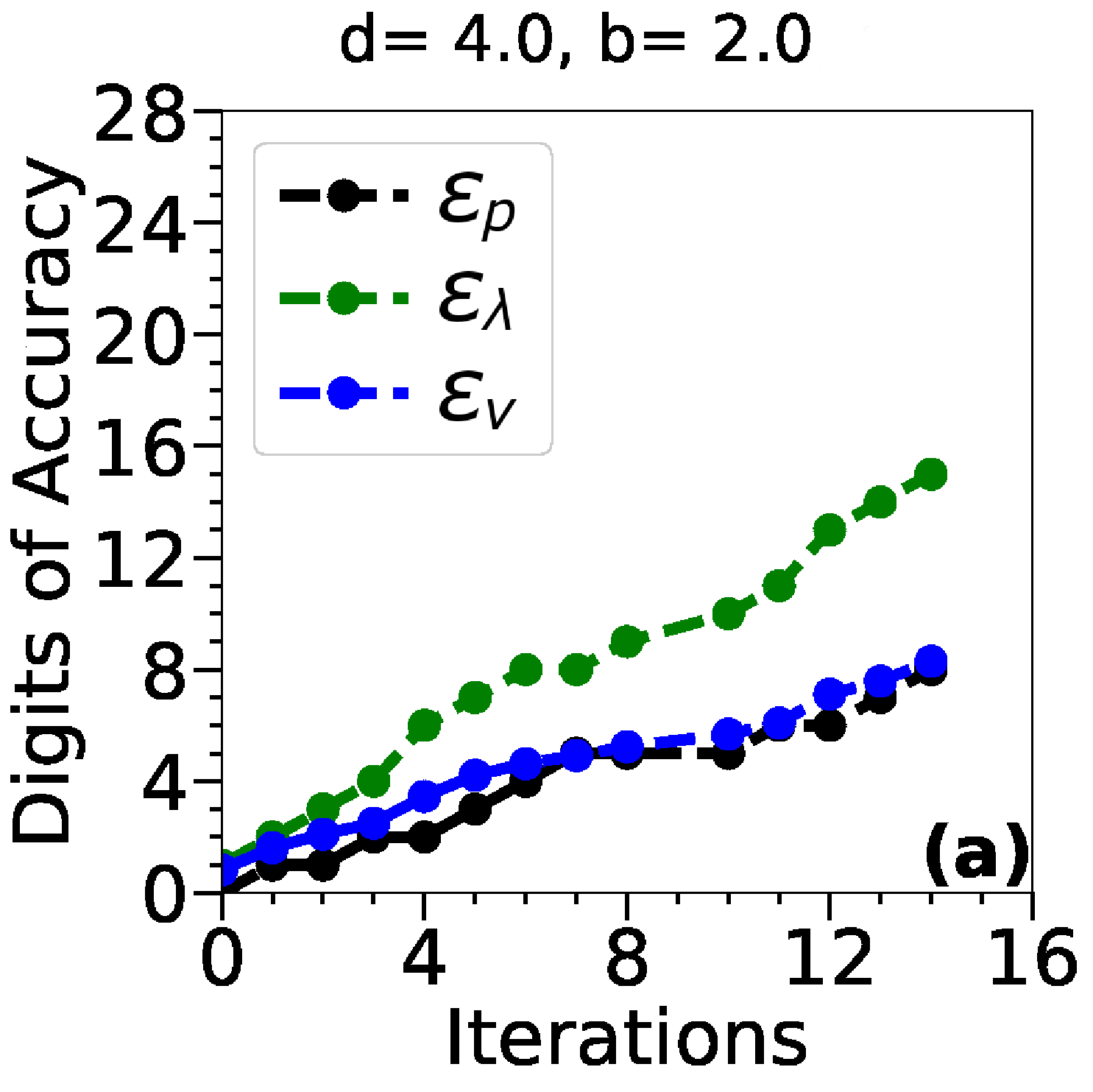}
}
\hfill
\subfloat{
    \includegraphics[width=0.32\textwidth]{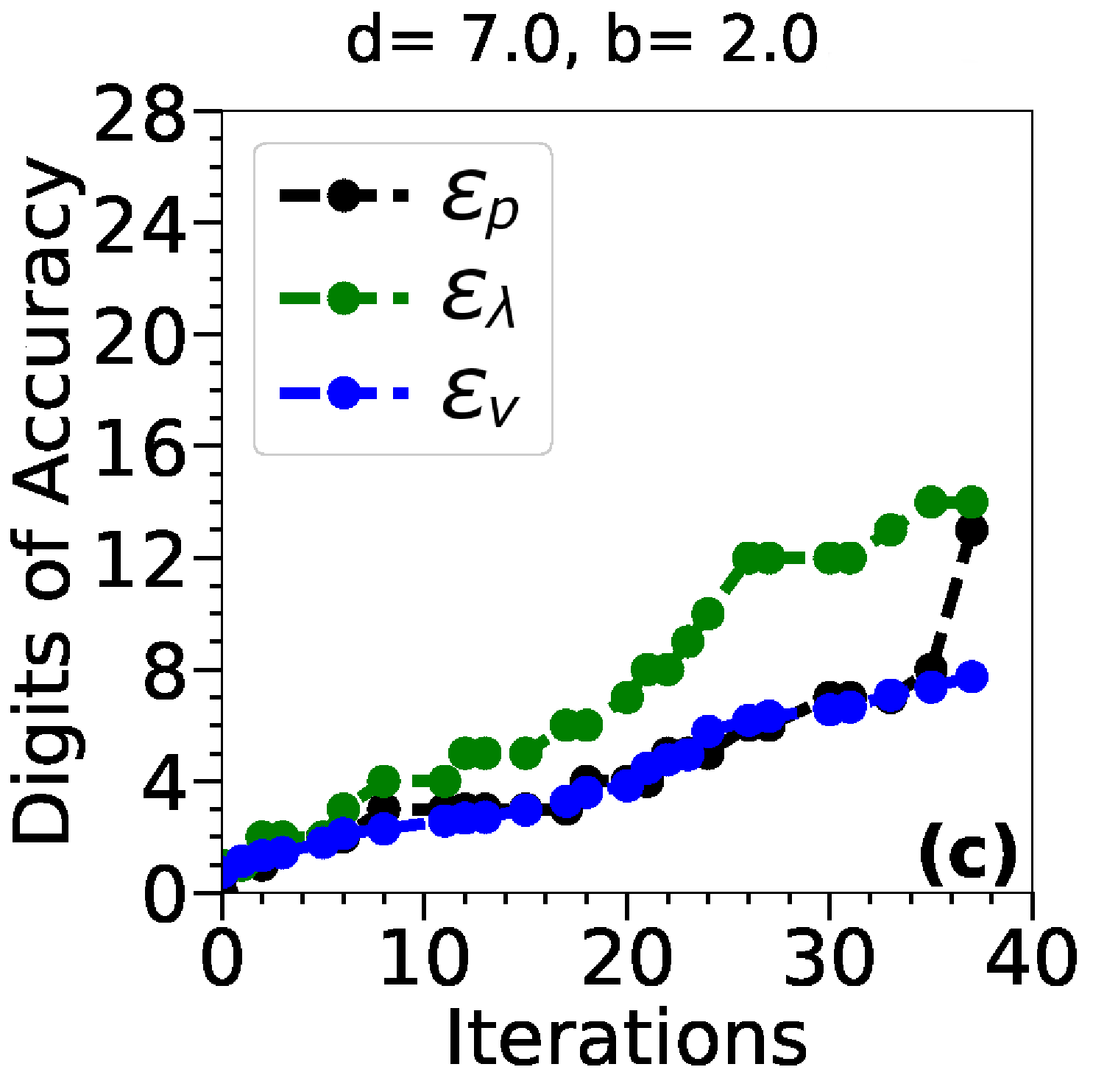}
}
\hfill
\subfloat{
    \includegraphics[width=0.32\textwidth]{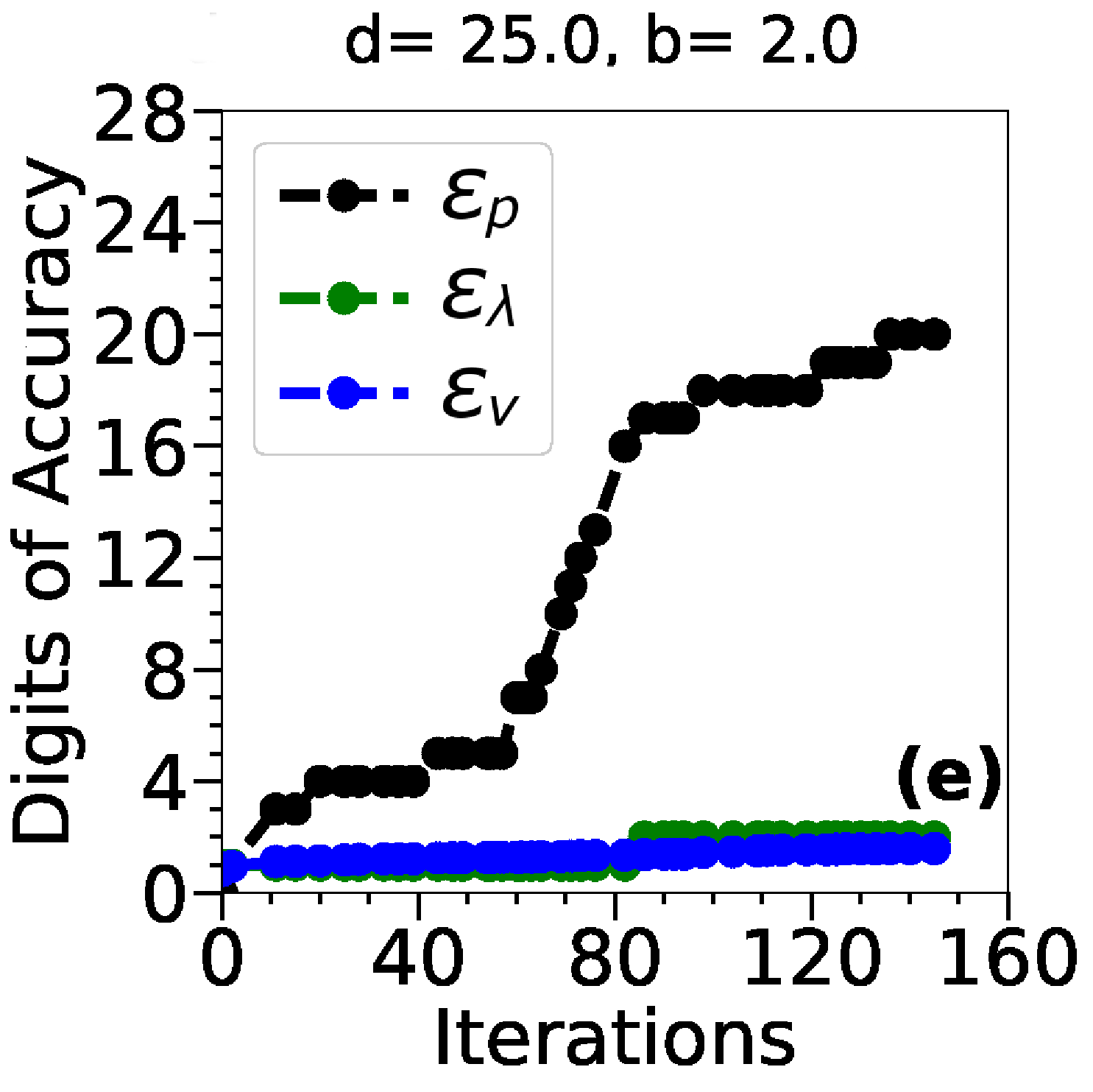}
}

\vspace{0.15cm}

\subfloat{
    \includegraphics[width=0.32\textwidth]{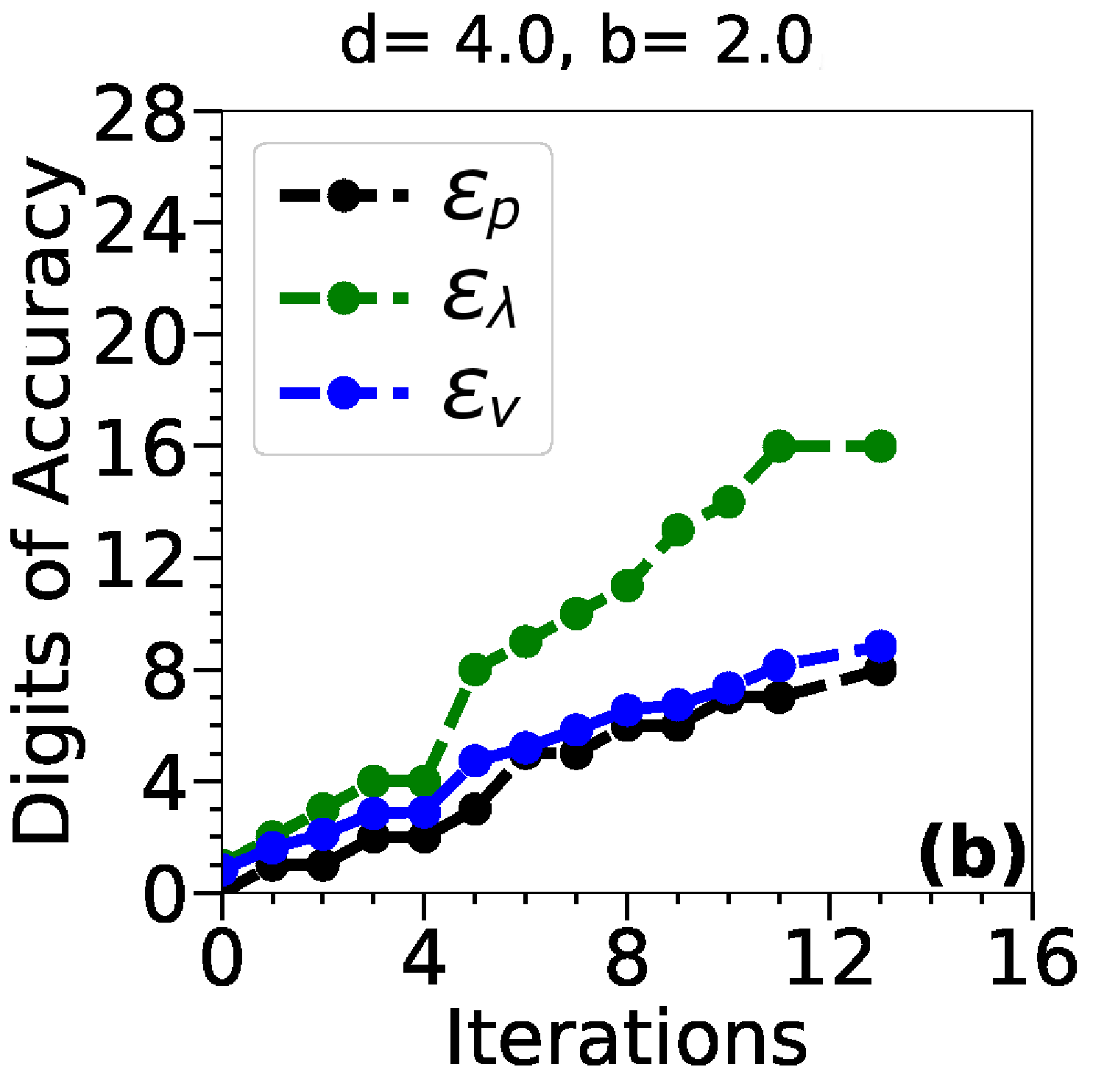}
}
\hfill
\subfloat{
    \includegraphics[width=0.32\textwidth]{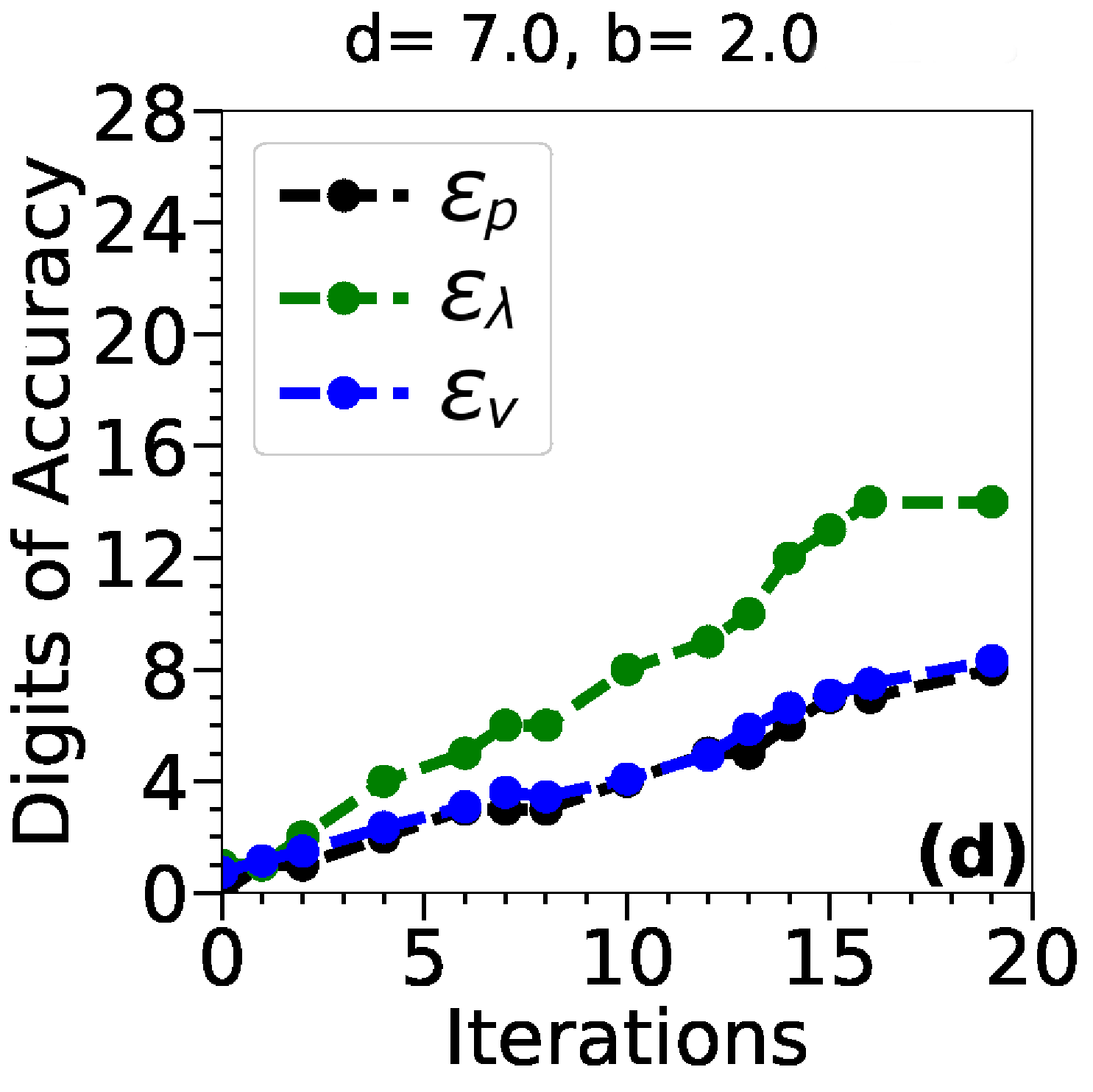}
}
\hfill
\subfloat{
    \includegraphics[width=0.32\textwidth]{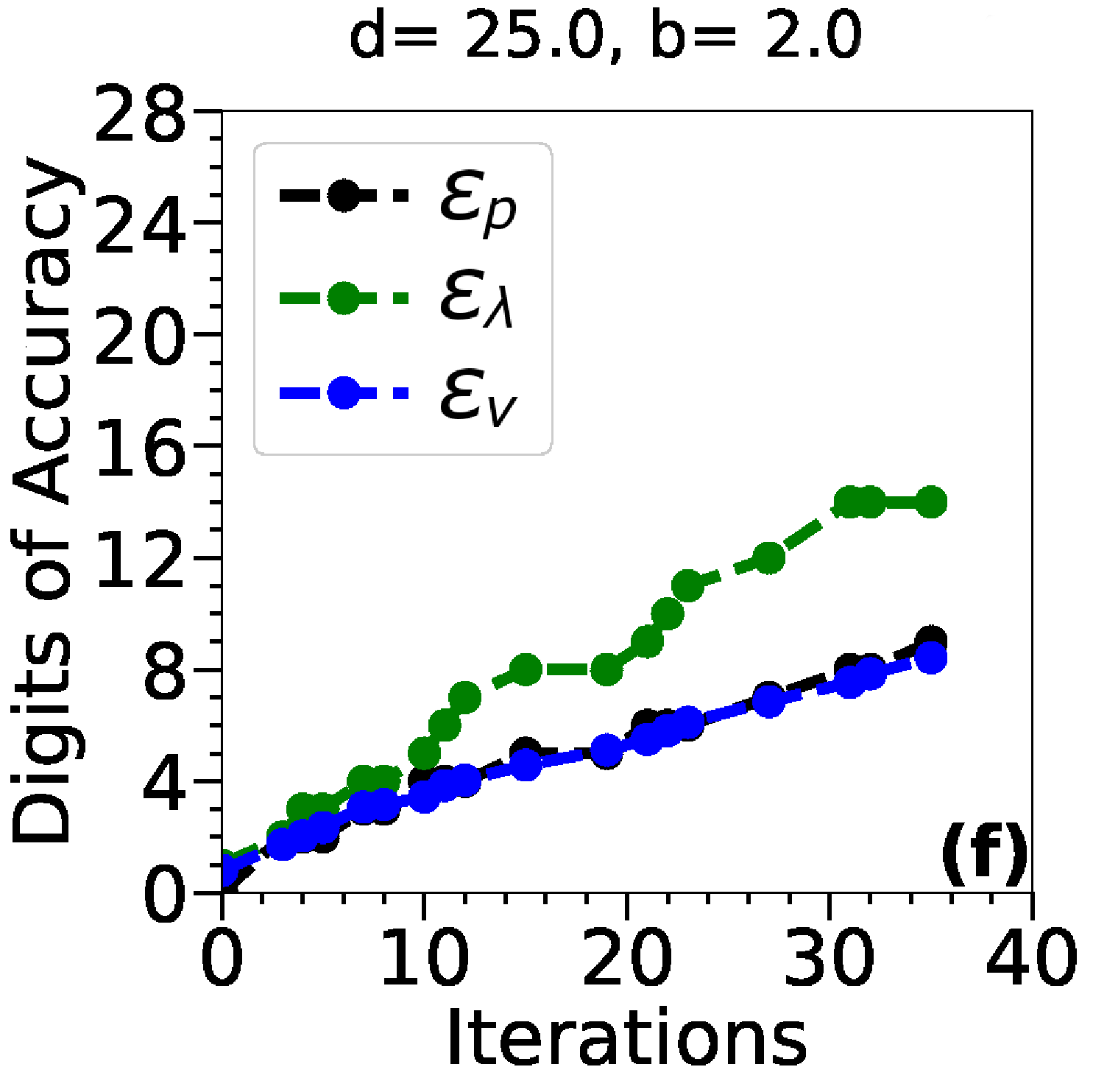}
}

\caption{Digits of accuracy of the lowest eigenvalue and of the corresponding eigenvector 
for the TDA Hamiltonian diagonalization with $J^\pi = 0^+$ and $b=2$, performed with 
SA (a,c,e) and with QA (b,d,f). The value $\varepsilon_p = -\log_{10}(\mathrm{precision})$ 
is also reported.}
\label{fig:dwave_real_b2}
\end{figure*}
Taking into account all the considerations discussed above, we formulate  Algorithm~\ref{full_spectrum_eigensolver_sevp} and Algorithm~\ref{full_spectrum_eigensolver_gevp}, which are designed to compute the complete eigenspectrum of the SEVP and the GEVP, respectively. 

A potential performance issue arises because computing the maximum magnitude requires solving two QUBO problems at each iteration, resulting in a total of \(2 \times N\) QUBO problems for \(N\) iterations.
To mitigate this computational cost, we implement a strategy where, at each iteration, the QUBO result that is not immediately used (i.e., the eigenpair corresponding to the minimum magnitude) is stored and reused in the subsequent iteration. This approach reduces the total number of QUBO problems to solve, thereby improving computational efficiency.

Specifically, for the SEVP, only \(N\) iterations are required if the matrix \(A\) is strictly positive definite or strictly negative definite, whereas \(N + 1\) iterations are needed if \(A\) is not strictly positive or negative definite. The same applies to the GEVP: only \(N\) iterations are needed if the matrix pair \((A, B)\) satisfies strict definiteness conditions; otherwise, \(N + 1\) iterations are necessary.

\begin{figure*}[t]
\centering

\begin{minipage}{0.32\textwidth}
\centering
\includegraphics[width=\linewidth]{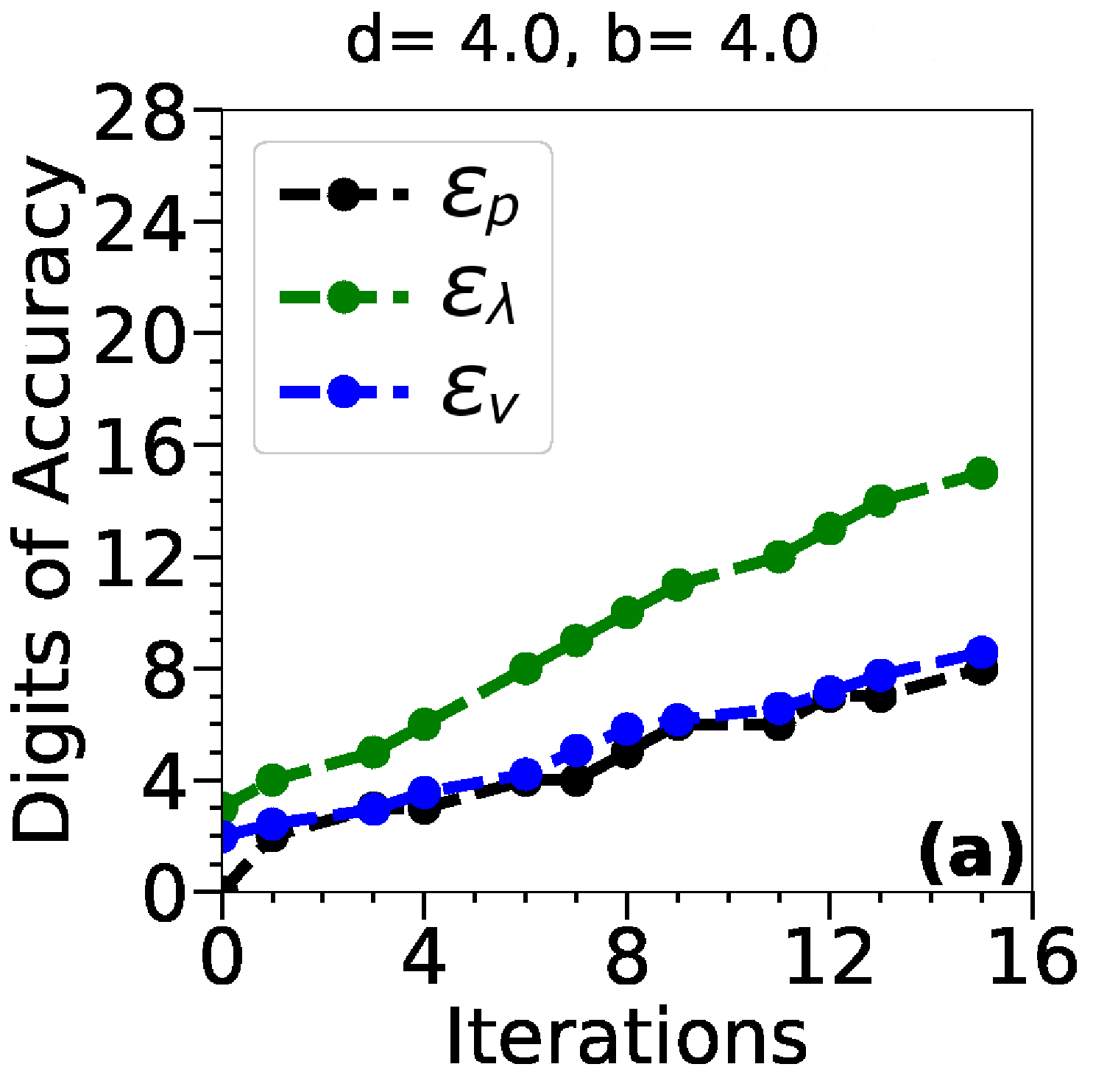}

\vspace{0.15cm}

\includegraphics[width=\linewidth]{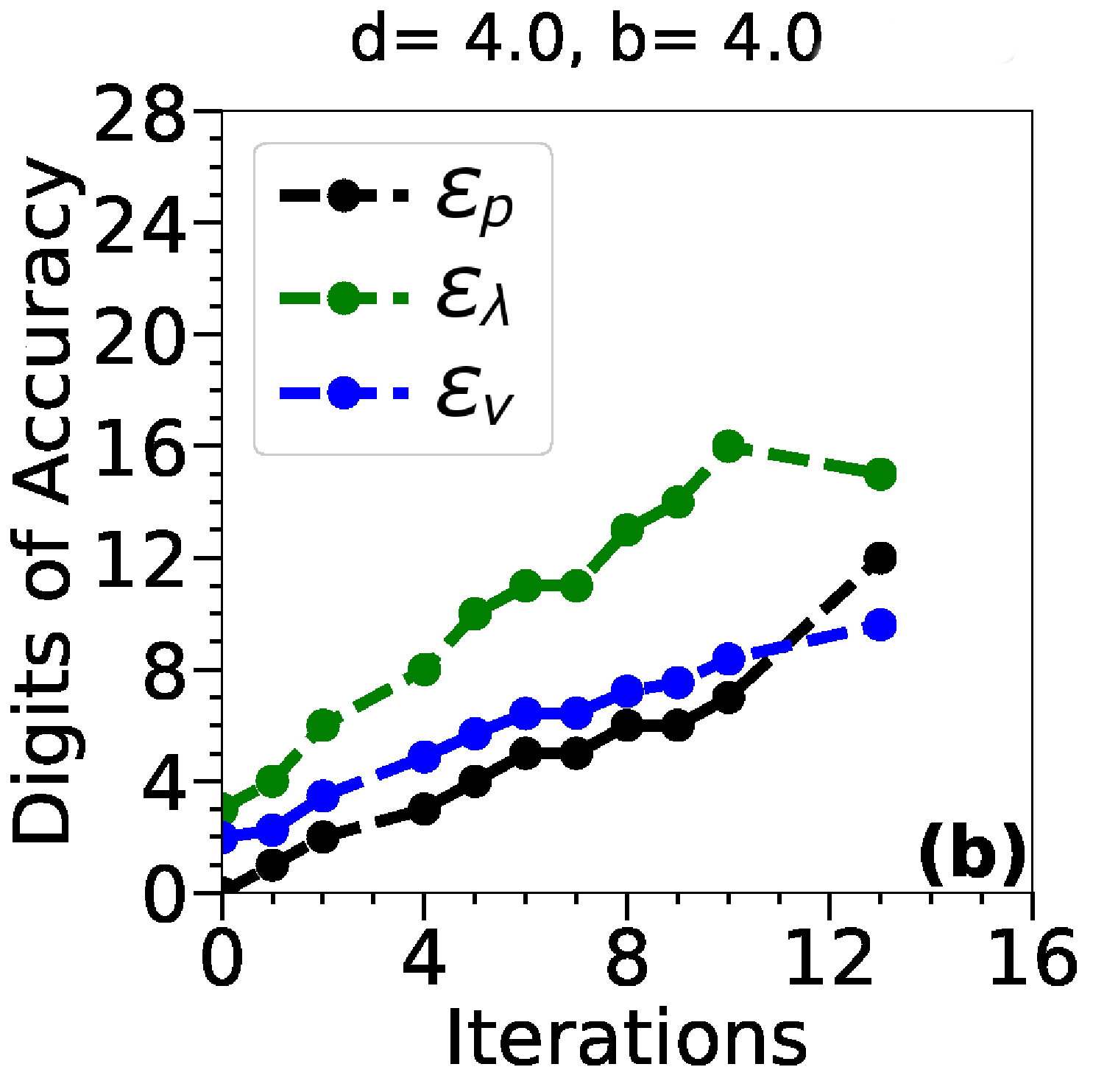}
\end{minipage}
\hfill
\begin{minipage}{0.32\textwidth}
\centering
\includegraphics[width=\linewidth]{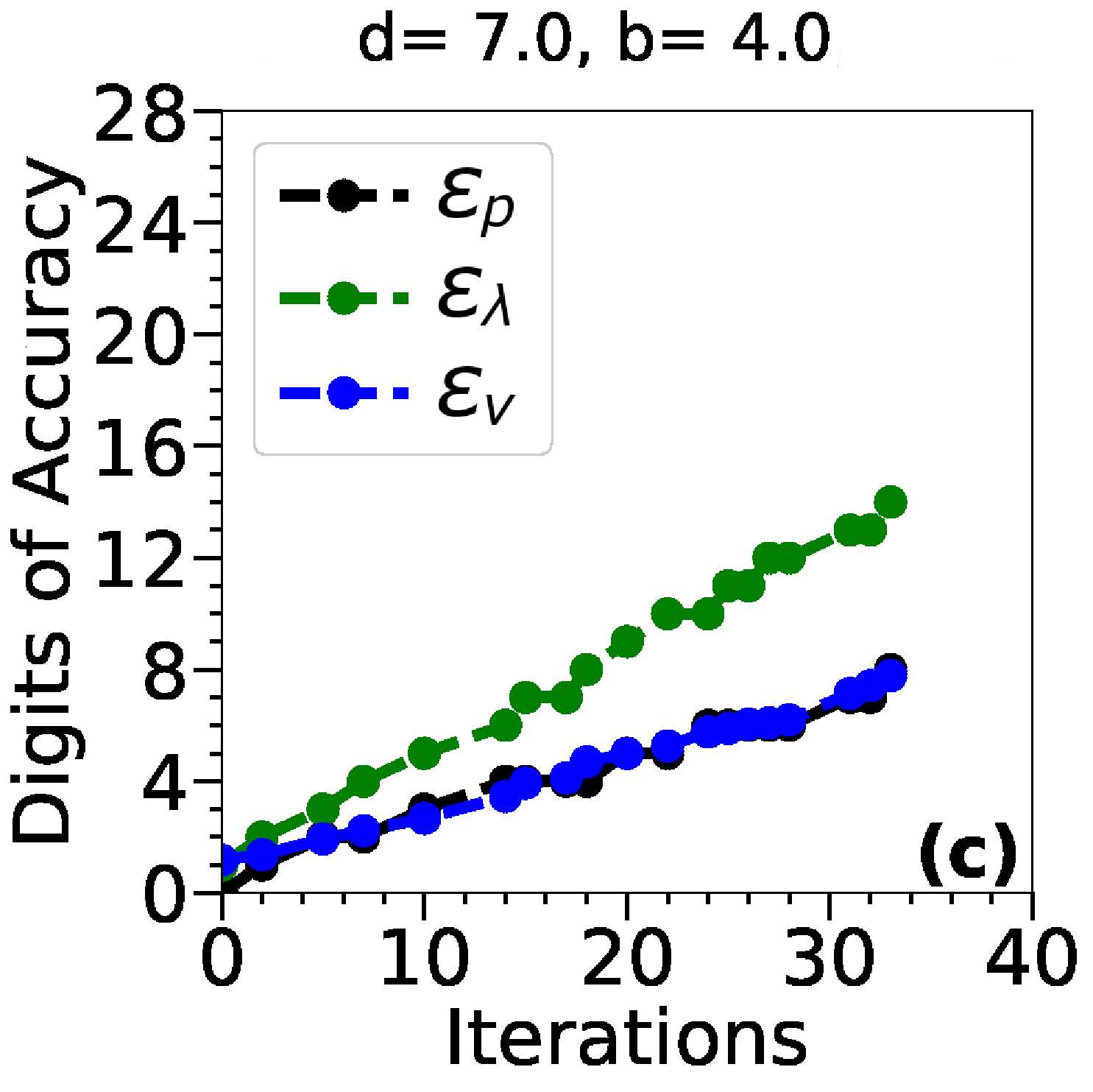}

\vspace{0.15cm}

\includegraphics[width=\linewidth]{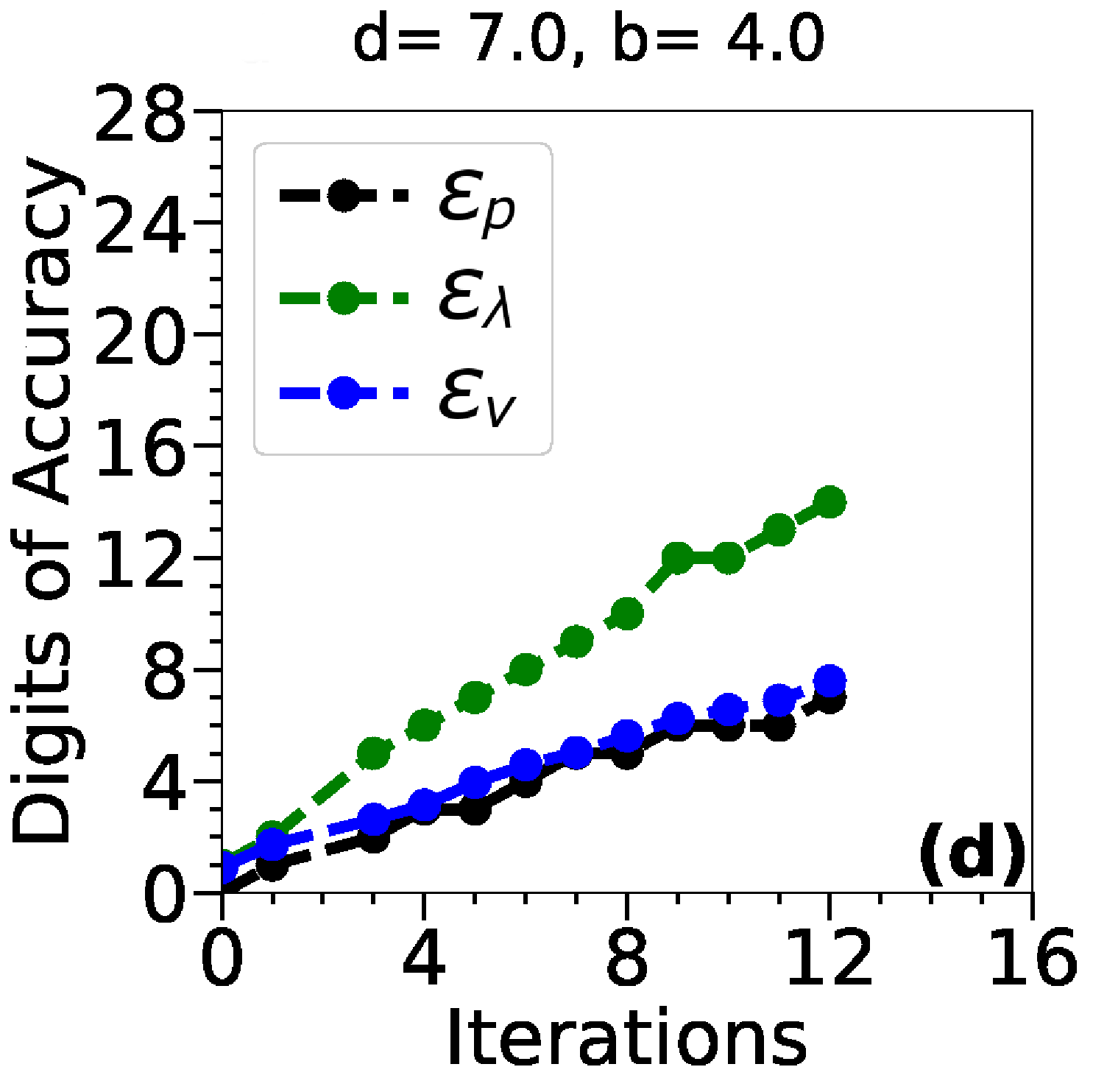}
\end{minipage}
\hfill
\begin{minipage}{0.32\textwidth}
\centering
\includegraphics[width=\linewidth]{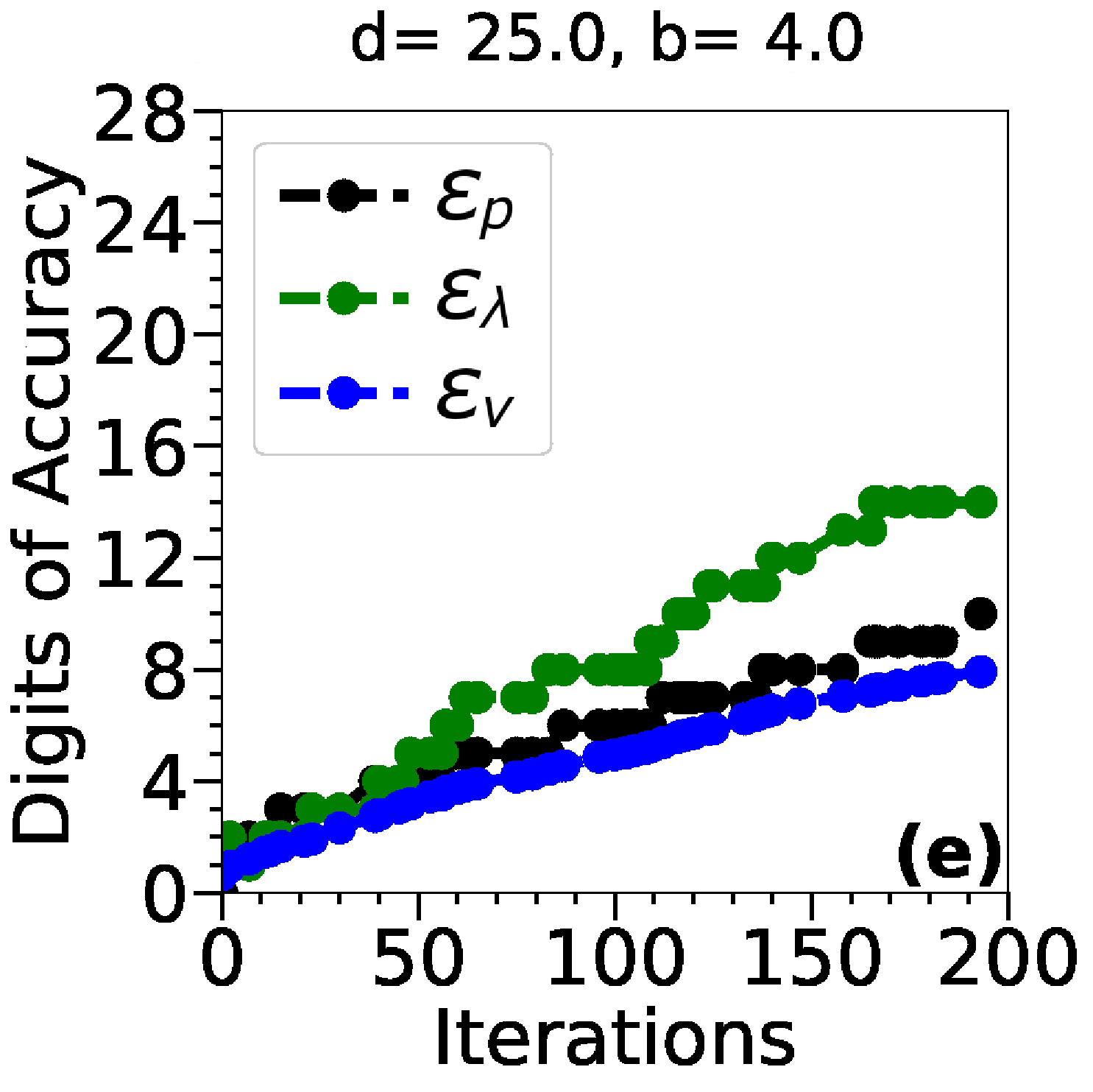}

\vspace{0.15cm}

\includegraphics[width=\linewidth]{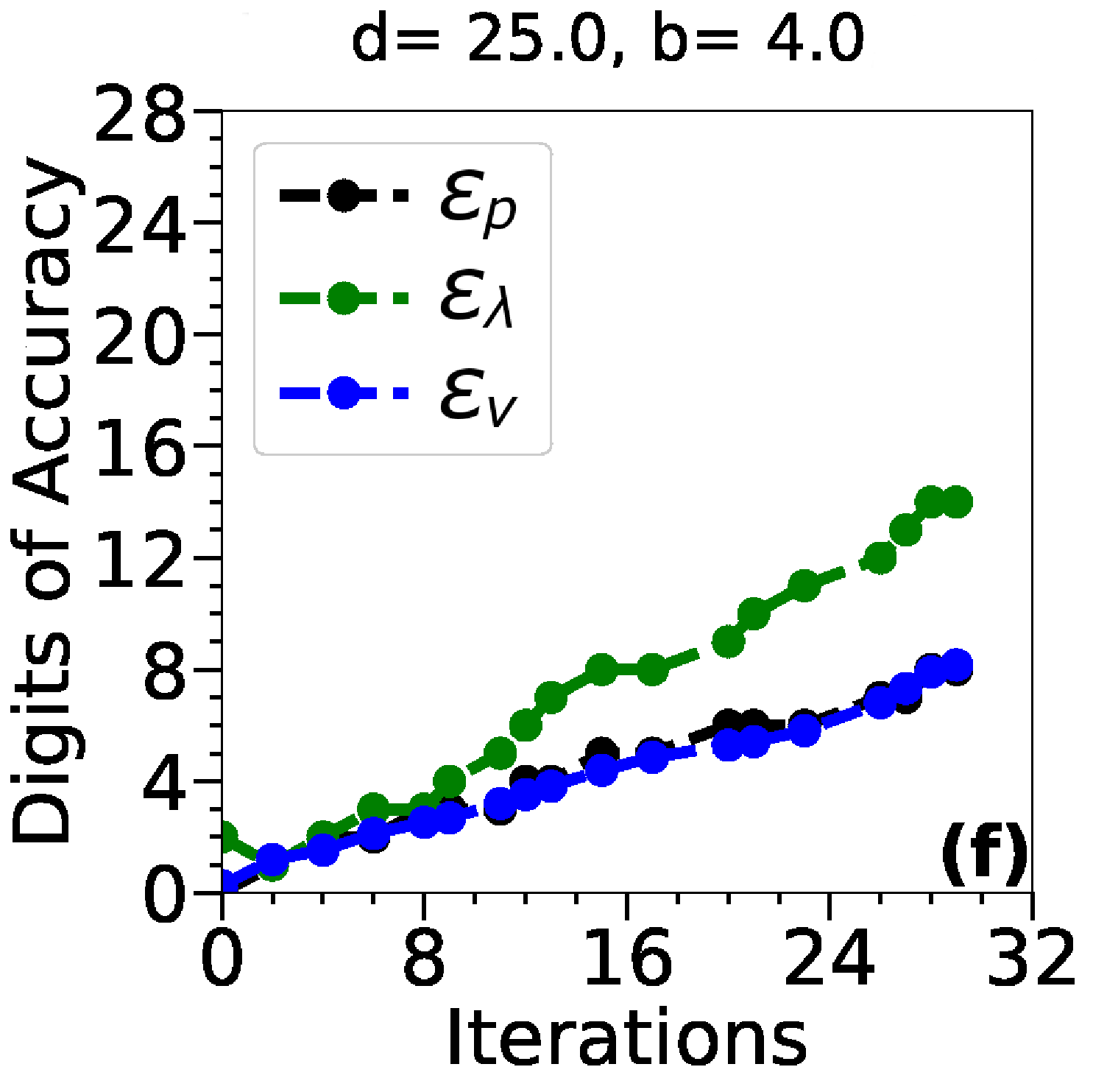}
\end{minipage}

\caption{Same as Fig.~\ref{fig:dwave_real_b2}, for $b=4$.}
\label{fig:dwave_real_b4}
\end{figure*}
Hereafter, we analyse the computational cost of each of the aforementioned deflation methods. All three methods share the same asymptotic complexity when extracting the full spectrum of dense matrices: 
\begin{itemize}
  \item \textbf{Total time:} $O(n^{3})$ for each method (arising from $O(n^{2})$ work per deflation step, repeated $\simeq n$ times; this excludes the cost of the eigensolver used to compute individual $(\lambda, v)$ pairs).
  \item \textbf{Memory:} $O(n^{2})$ (storage of $A$, $B$, and dense auxiliary matrices).
\end{itemize}
Despite sharing the same asymptotic cost, the methods differ in numerical robustness, hidden constants, and structure preservation. 

The Hotelling method computes $u = Bv$ and performs a rank-one correction of $A$ at each deflation step, for a cost of $O(n^{2})$. It is simple and has low hidden constants, but may introduce numerical instabilities unless combined with re-orthogonalization or filtering. 

The Orthogonal Projection approach also performs $u = Bv$, followed by orthogonalization in the $B$-inner product and projection, again yielding an $O(n^{2})$ cost per step. Its cost is stable across iterations, and it is typically the most numerically robust method, producing fewer spurious eigenpairs. 

Finally, the Householder deflation  applies congruent reflectors to progressively smaller active submatrices. The $i$-th iteration costs $O(m \cdot n)$ with $m = n - i$, leading to a total of $O(n^{3})$ but with smaller leading constants because later steps are cheaper. 

This approach preserves symmetry and congruence and is often efficient in practice on dense problems, though repeated updates of $A$ and $B$ may compromise the orthogonality of the computed eigenvectors.

\section{Numerical implementation and results}
As a study case, we consider $^4$He. The starting Hamiltonian is defined as

\begin{equation}\label{eq: intrinsic hamiltonian}
H = T_{\rm int} + V_D.
\end{equation}

Here, $T_{\rm int}$ denotes the intrinsic kinetic energy, and $V_D$ is the Daejeon interaction derived from the \textit{NN} component of the N3LO potential \cite{Shirok2016} through a two-step procedure. First, the \textit{NN} potential is softened using the SRG method \cite{SRG} with flow parameter $\lambda = 1.5~\mathrm{fm}^{-1}$ then, a phase-equivalent transformation is applied to determine an optimal parametrization of the \textit{NN} force.
 We use this Hamiltonian to generate a HF basis encompassing 3 major shells. We use the HF states to create the TD phonon basis and generate the 2-phonon basis by deriving and solving iteratively the EMPM Eq. \eqref{eq:EMPM};  the basis so constructed is adopted to solve the final eigenvalue problem in the multiphonon space Eq. \eqref{eq:eigfull}.

\begin{figure*}[t]
\centering

\begin{minipage}{0.32\textwidth}
\centering
\includegraphics[width=\linewidth]{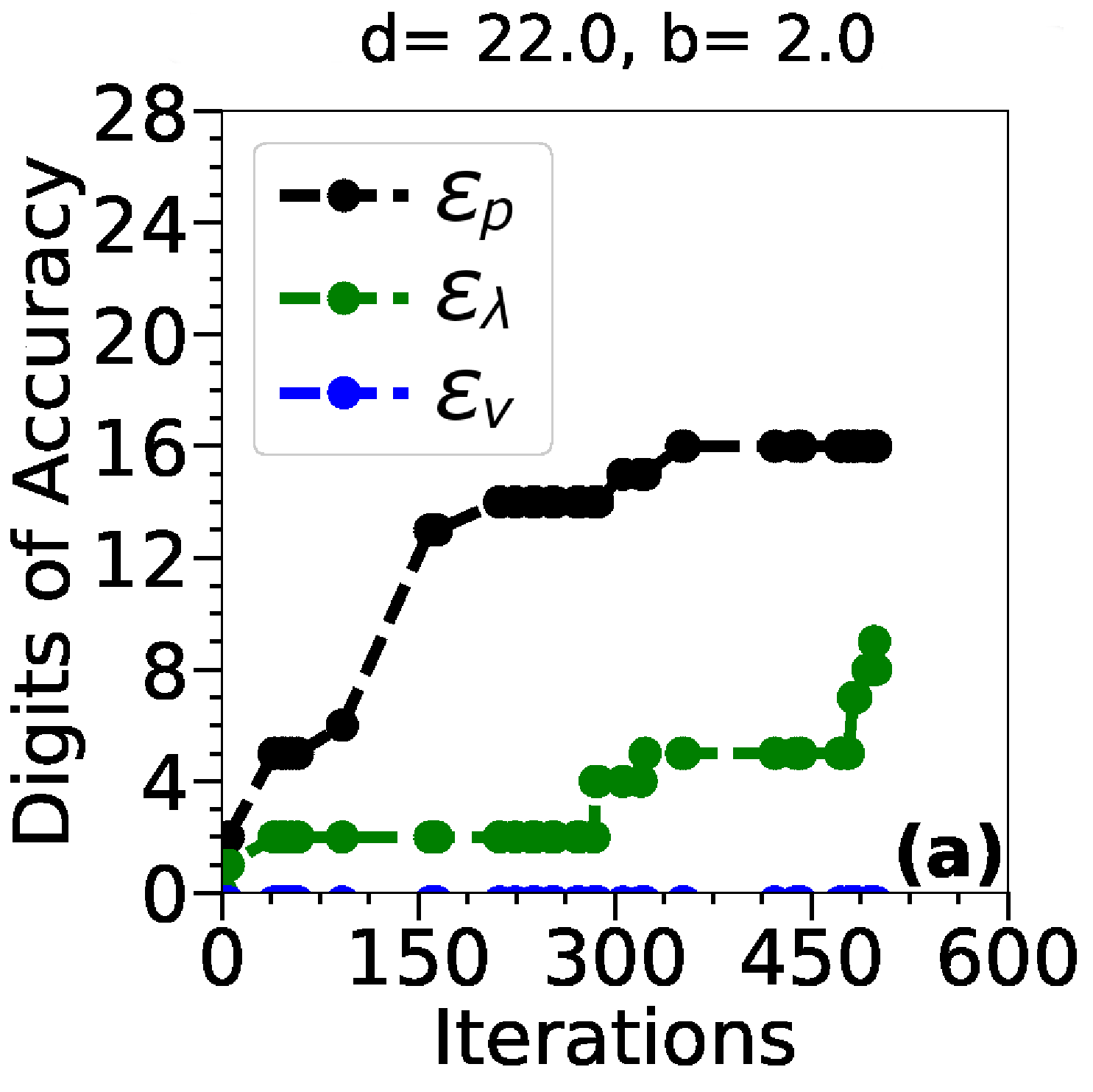}

\vspace{0.15cm}

\includegraphics[width=\linewidth]{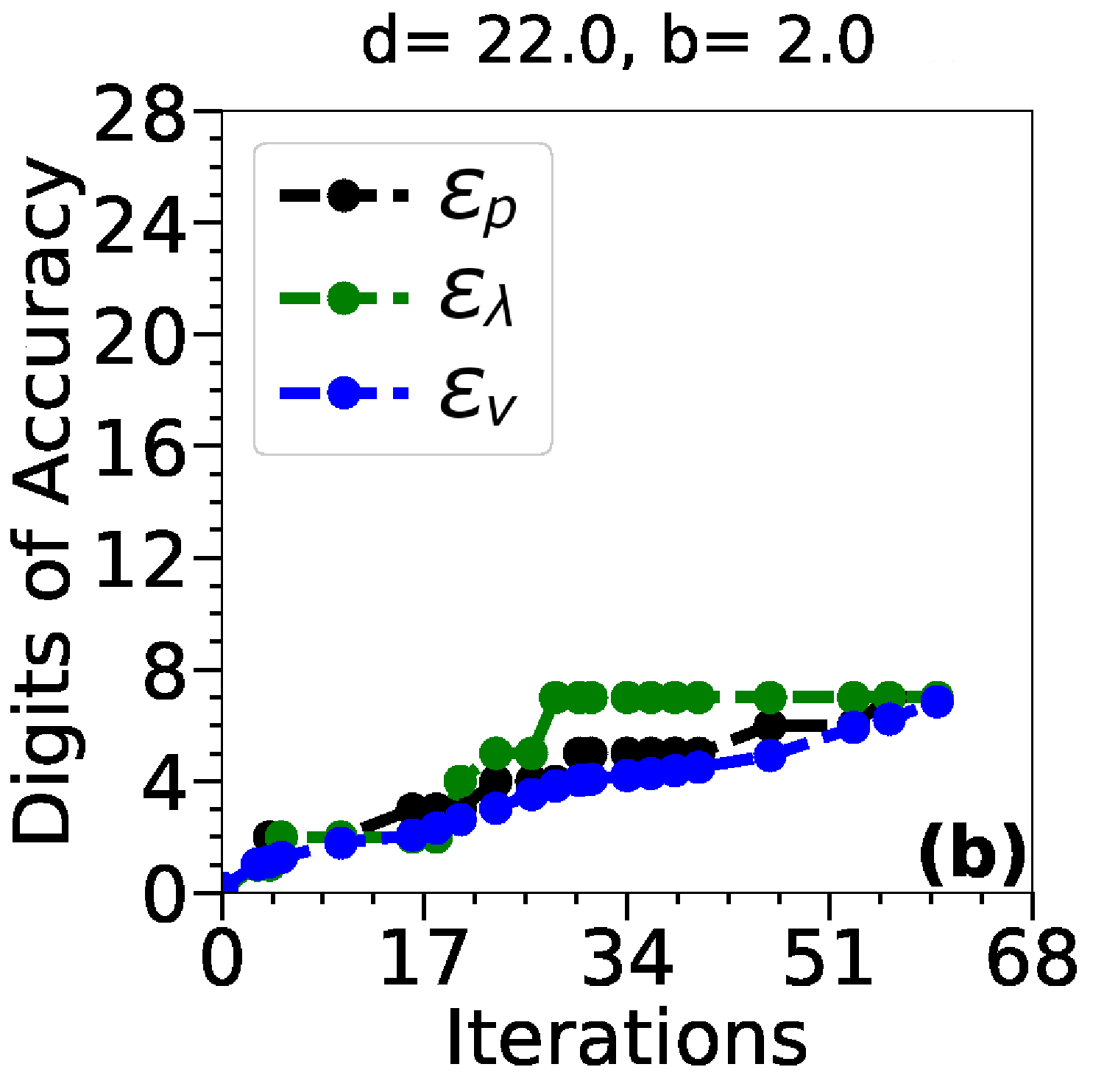}
\end{minipage}
\hfill
\begin{minipage}{0.32\textwidth}
\centering
\includegraphics[width=\linewidth]{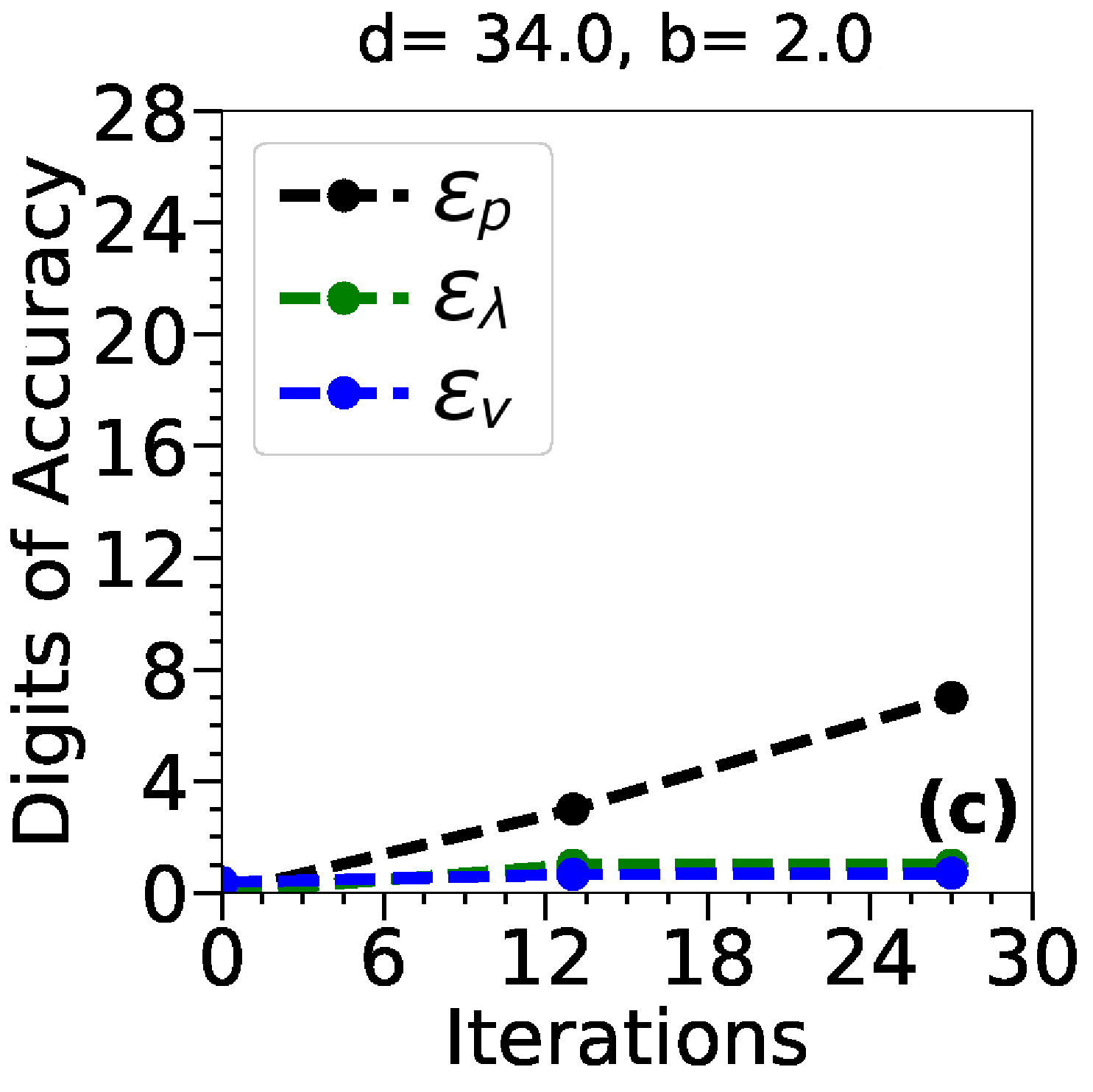}

\vspace{0.15cm}

\includegraphics[width=\linewidth]{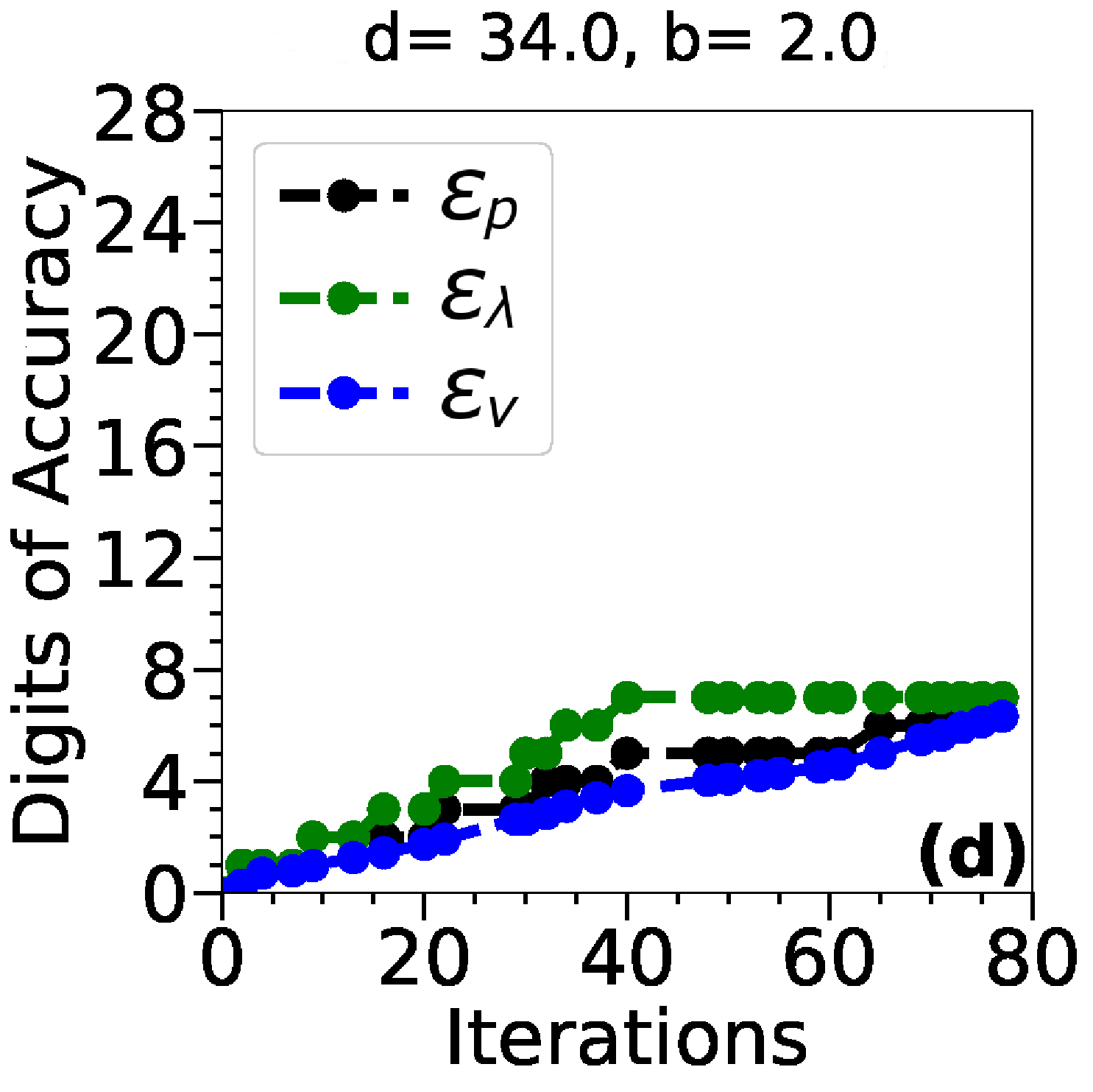}
\end{minipage}
\hfill
\begin{minipage}{0.32\textwidth}
\centering
\includegraphics[width=\linewidth]{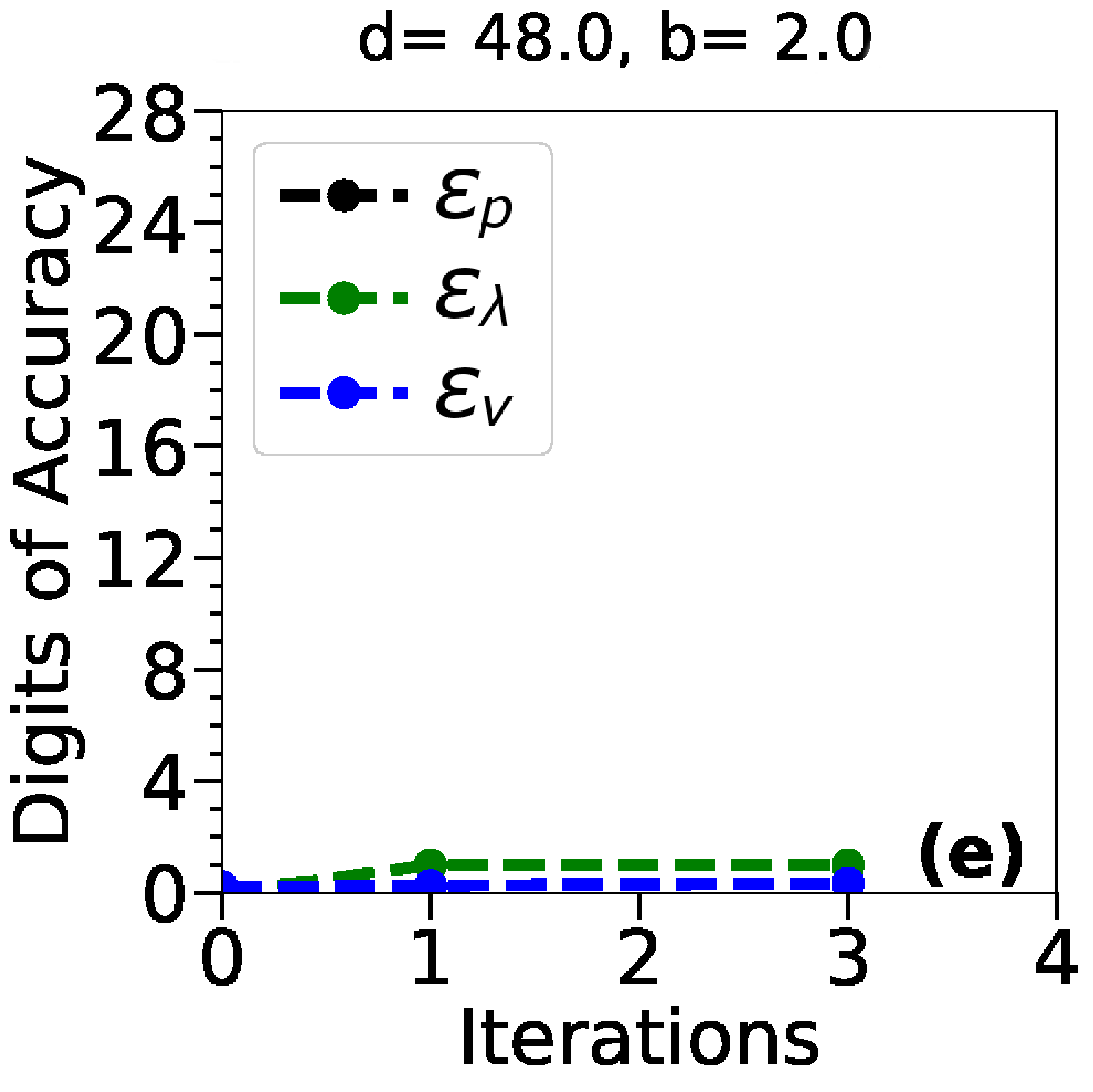}

\vspace{0.15cm}

\includegraphics[width=\linewidth]{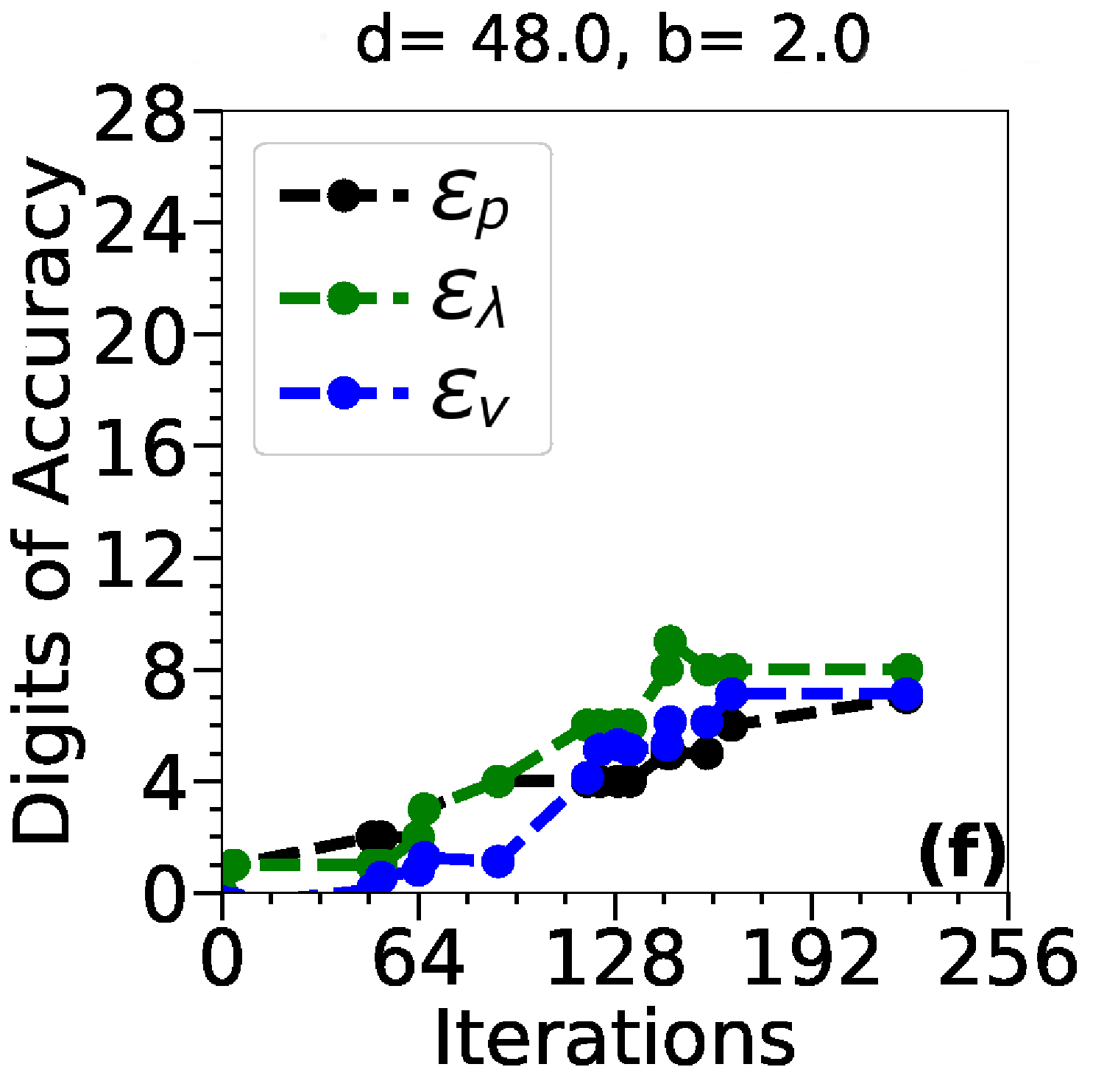}
\end{minipage}

\caption{Results for the solution of the GEVP within the Equation-of-Motion Phonon Method
with $J^\pi = 0^+,1^-,2^+$ and $b=2$.}
\label{fig:genEIG_b2}
\end{figure*}

\begin{figure*}[t]
\centering

\begin{minipage}{0.32\textwidth}
\centering
\includegraphics[width=\linewidth]{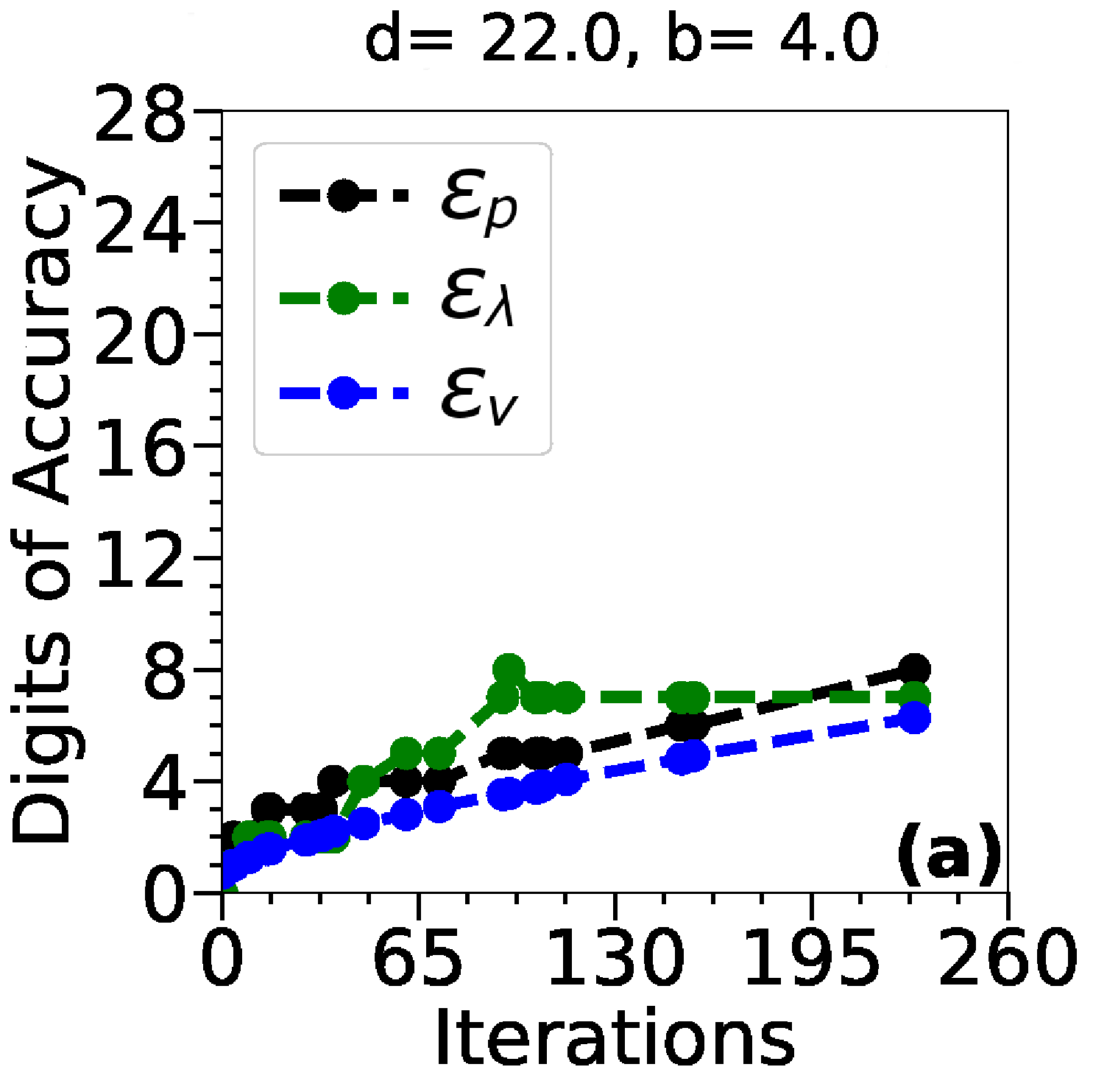}

\vspace{0.15cm}

\includegraphics[width=\linewidth]{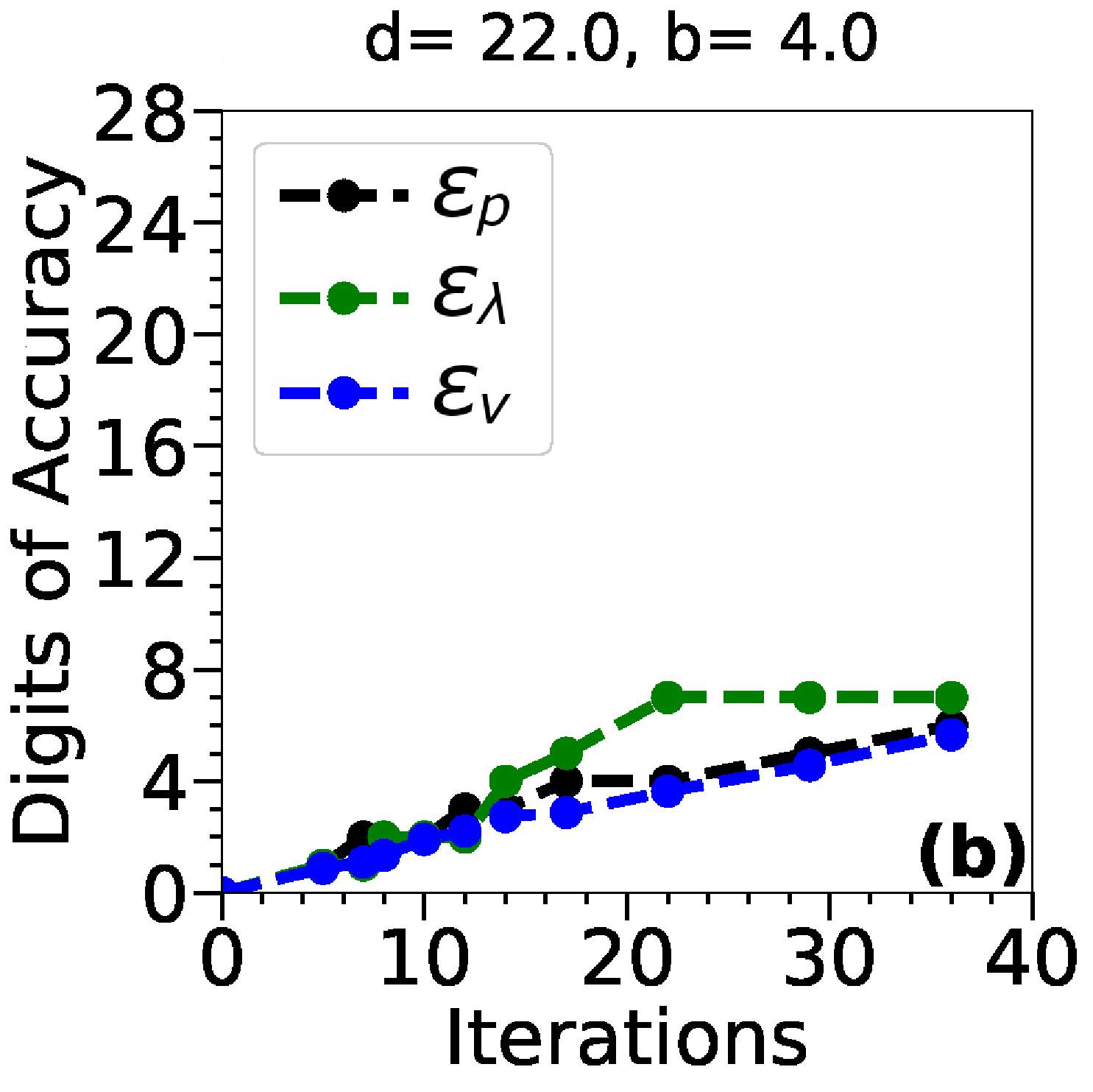}
\end{minipage}
\hfill
\begin{minipage}{0.32\textwidth}
\centering
\includegraphics[width=\linewidth]{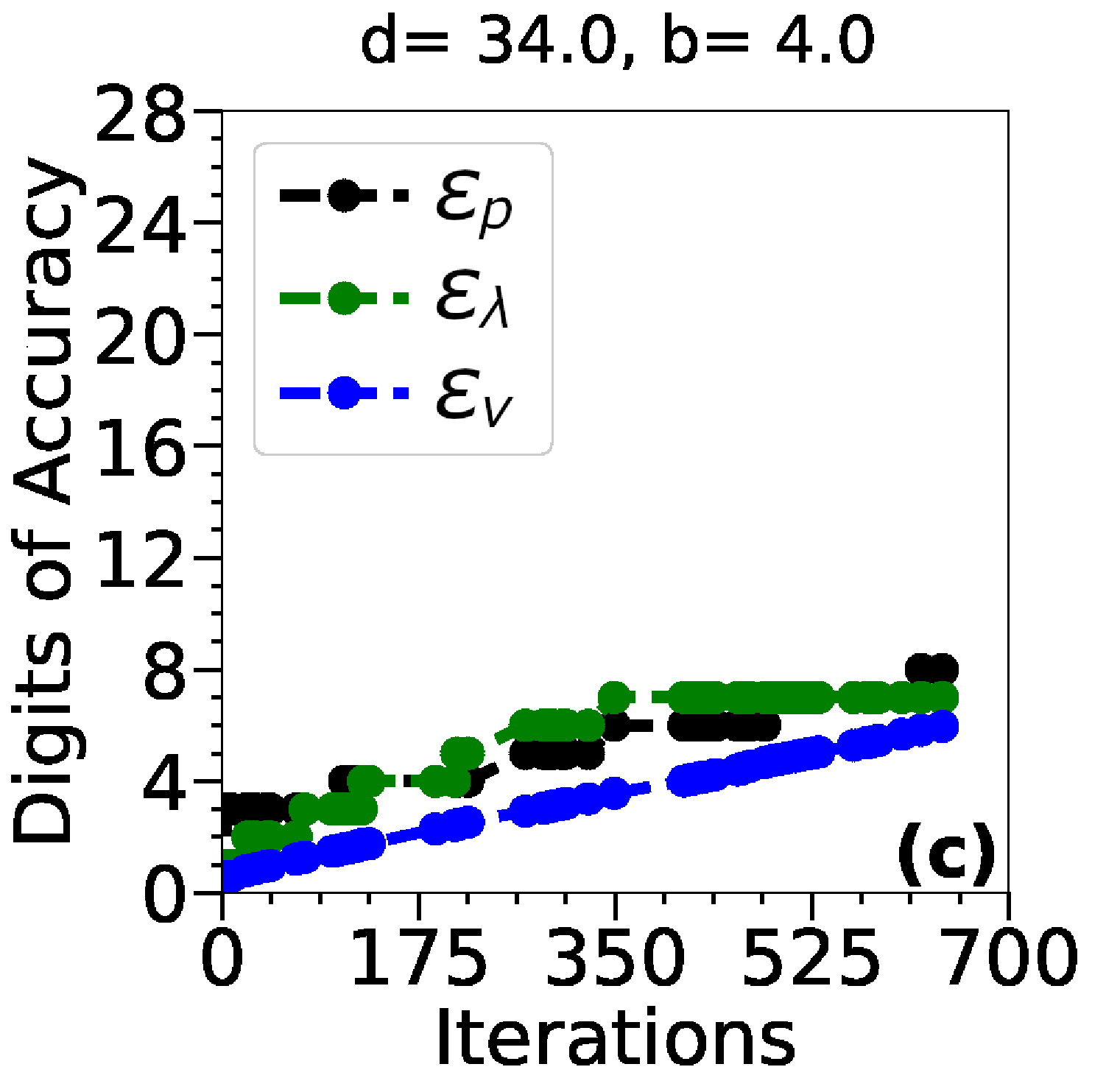}

\vspace{0.15cm}

\includegraphics[width=\linewidth]{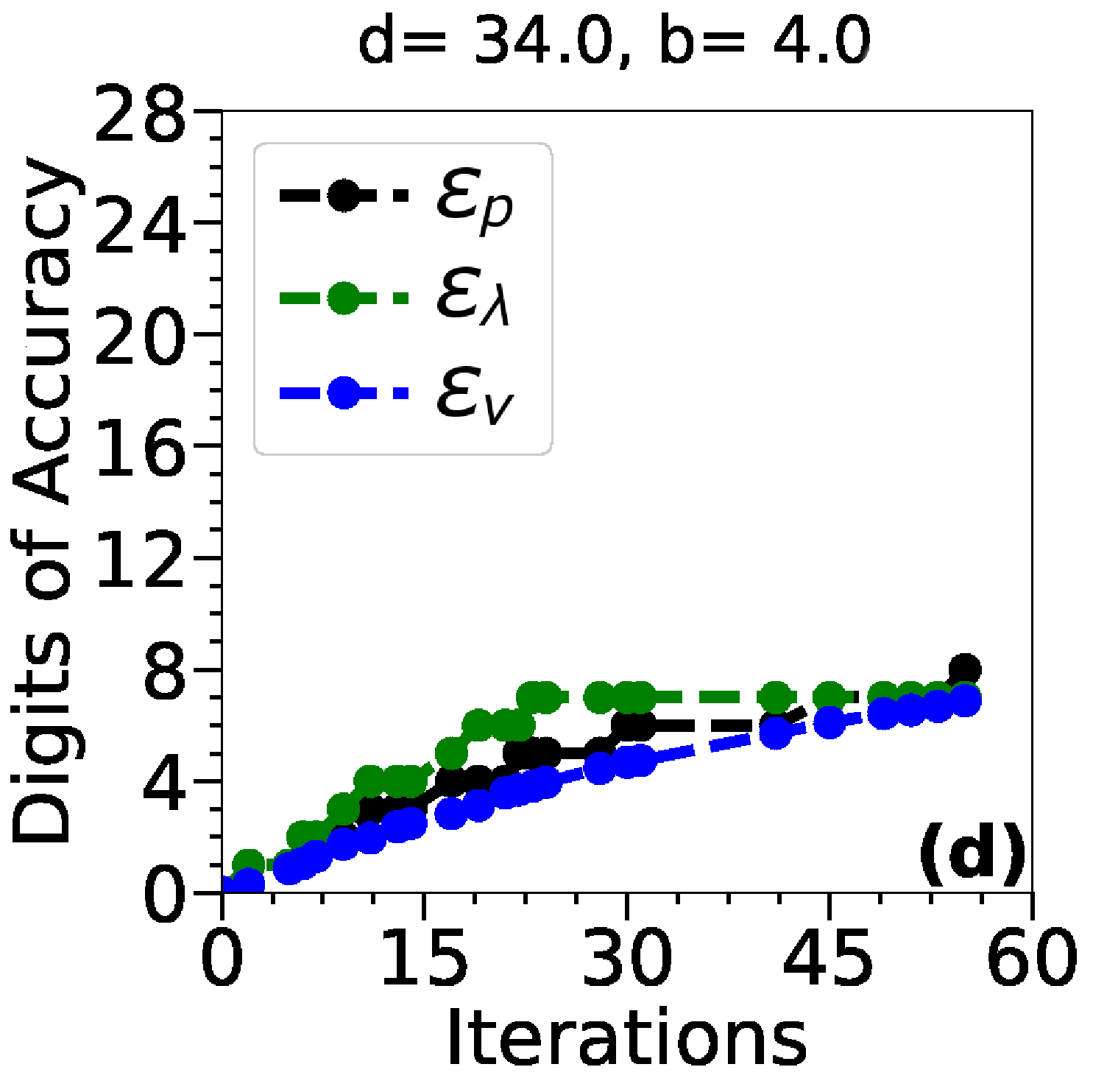}
\end{minipage}
\hfill
\begin{minipage}{0.32\textwidth}
\centering
\includegraphics[width=\linewidth]{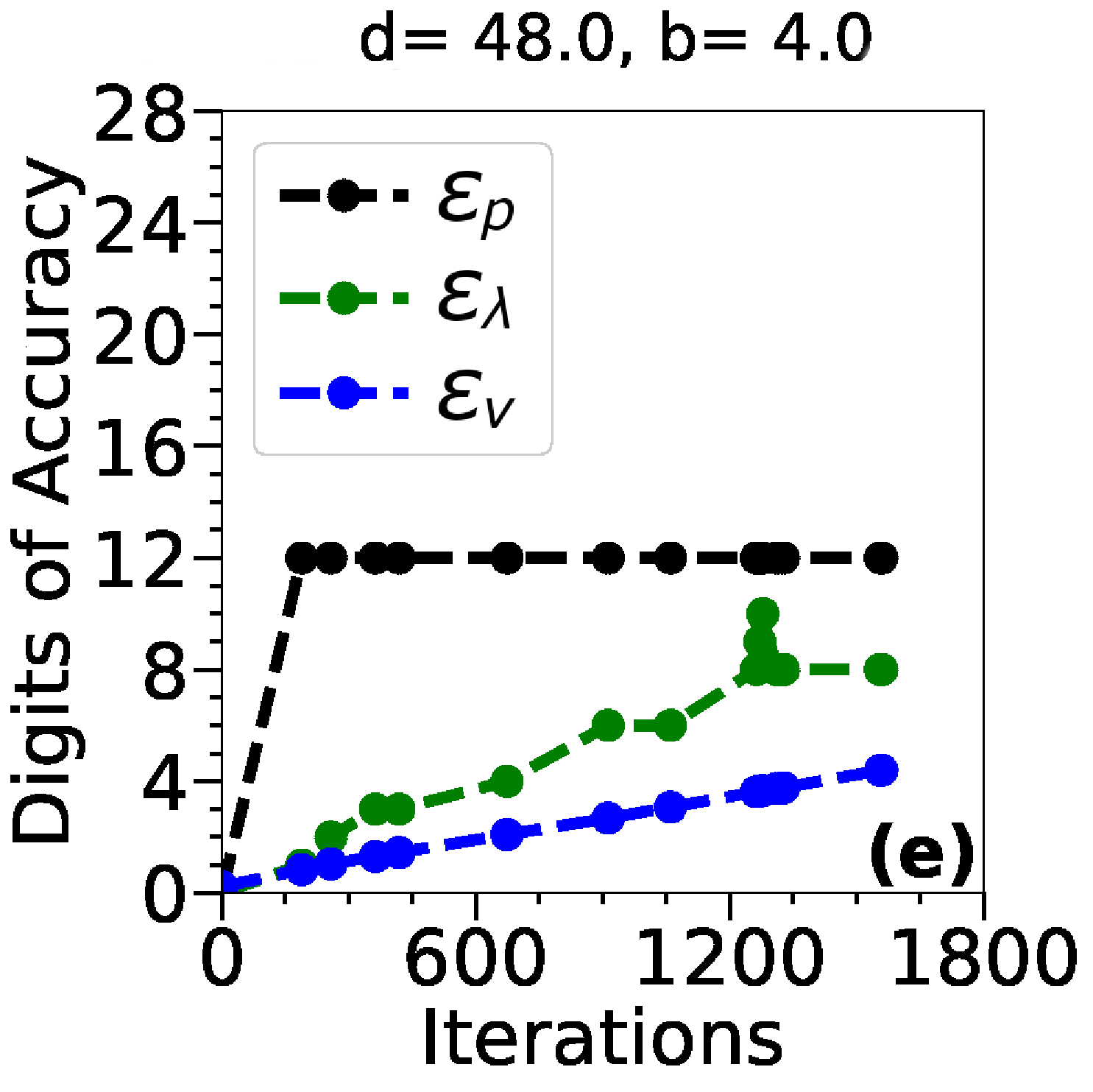}

\vspace{0.15cm}

\includegraphics[width=\linewidth]{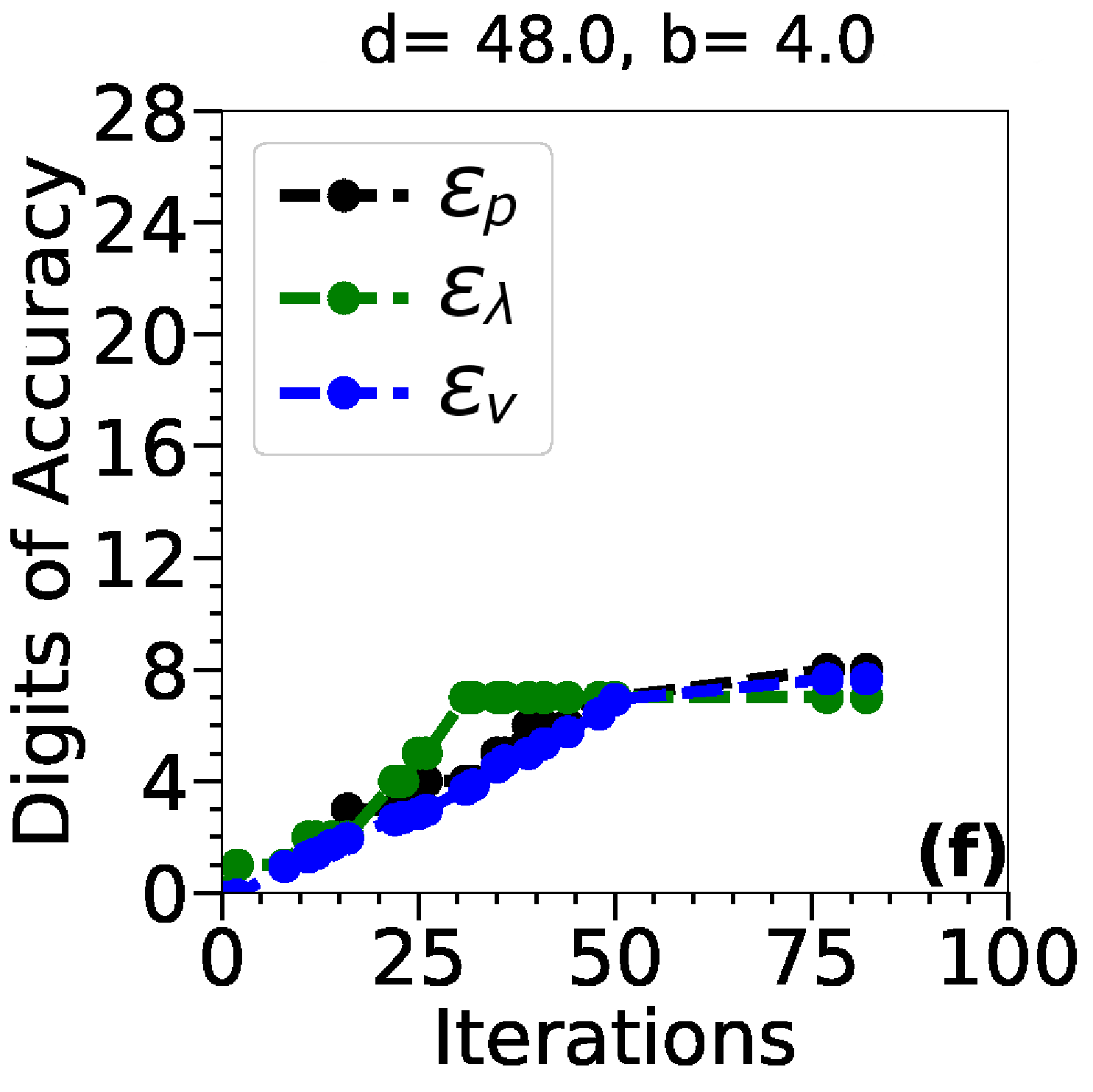}
\end{minipage}

\caption{Same as Fig.~\ref{fig:genEIG_b2}, for $b=4$.}
\label{fig:genEIG_b4}
\end{figure*}
\subsection{Ground state}
We begin by evaluating the performance of the QUBO-based algorithm for computing the ground-state eigenpair using a D-Wave quantum annealer. All quantum annealing (QA) calculations were carried out on the Advantage System 6.4 \cite{DWaveSystems}, which provides 5,614 working qubits. For comparison, we also performed the same simulations using a classical Simulated Annealing (SA) solver.

The analysis covers different multipolarities, corresponding to different matrix dimensions $d$, and two bit resolutions, $b = 2$ and $b = 4$. The starting point in the plots corresponds to the output of the initial-guess phase, where the precision parameter $p$ (see Section~\ref{sec:Qubo_based_algorithm}) is set to $1$. The subsequent evolution shows how the accuracy improves during the descent phase as the discretized search cube is progressively refined by increasing $p$.

For each configuration, we monitor the errors on the eigenvalue ($eval\_error$) and eigenvector ($evec\_error$) as functions of both $p$ and of the iteration number ($N_{it}$) . Here, $eval\_error$ denotes the deviation of the computed eigenvalue from the exact one, while $evec\_error$ is the Euclidean distance between the computed eigenvector and the true ground-state eigenvector. Reference solutions were obtained using standard numerical eigensolvers.

We first discuss the results shown in Fig.~\ref{fig:dwave_real_b2} for $b = 2$. The figure reports the number of correct digits of the lowest eigenvalue and the corresponding eigenvector for the TDA Hamiltonian, comparing SA and QA. Panels (a), (c), and (e) refer to SA, while panels (b), (d), and (f) refer to QA. The parameter $\varepsilon_p = -\log_{10}(\text{precision})$ is also indicated. As test cases, we consider three Hamiltonians: the TDA Hamiltonian for $J^\pi = 0^+$ with dimension $d = 4$, the TDA Hamiltonian for $J^\pi = 1^-$ with dimension $d = 7$, and the full Hamiltonian for $J^\pi = 0^+$ with dimension $d = 25$.

Let us start with $J^\pi = 0^+$ (panels (a),(b)). We observe no significant difference between SA and QA in terms of the number of iterations ($N_{it} \simeq 14$).

As the dimension increases with $J^\pi = 1^-$ (panels (c), (d)), QA (panel (d)) begins to show better performance compared to SA (panel (c)), with the number of iterations decreasing from $N_{it} \simeq 40$ with SA to $N_{it} \simeq 12$ with QA.

The most interesting scenario is the full Hamiltonian case (panels (e), (f)). Here, SA (panel (e)) fails to improve the accuracy of the lowest energy eigenvalue beyond the initial guess phase. On the other hand, QA (panel (f)) achieves machine-precision accuracy on the computed eigenvalue within $N_{it} \simeq 30$ iterations.

We now analyse the case with increased bit resolution, setting $b=4$ (Fig.~\ref{fig:dwave_real_b4}). Compared to the $b=2$ case, we observe a significant improvement in the performance of QA over SA. For the largest matrix size $d=25$, SA (panel (e)) requires approximately $N_{it} \simeq 200$ iterations 
to achieve machine-precision accuracy on the lowest eigenvalue, whereas QA (panel (f)) reaches the same level of accuracy within only $N_{it} \simeq 30$ iterations.

Let's conclude this Section with an analysis of QA performance in solving a GEVP. Specifically, we discuss the results obtained for the EMPM Hamiltonian with $N=2$ and $J^\pi = 0^+, 1^-$ and $2^+$ of dimension $d=22, 34, 48$, respectively.

Even in the GEVP case, we find a meaningful improvement in performance when using QA over SA. This is already clear for $b=2$ (Fig. \ref{fig:genEIG_b2}), where using QA we get the eigenvalue with accuracy $10^{-8}$ in $N_{it} \simeq 30$, while with SA we need $N_{it} \simeq 450$, as shown in Fig. \ref{fig:genEIG_b2}(a),(b) for $J^\pi = 0^+$. For the eigenvector, the level of accuracy is $10^{-7}$ for $N_{it} \simeq 60$ with QA, versus $10^{-7}$ for $N_{it} \simeq 500$ with SA.
For $J^\pi = 1^-$ (Fig. \ref{fig:genEIG_b2}), we can formulate similar observations when comparing the result for SA, in Fig. \ref{fig:genEIG_b2}(c), with the one for QA in Fig. \ref{fig:genEIG_b2}(d). The SA is not sufficient to achieve 8-digit accuracy, while using QA the algorithm is able to gain $10^{-8}$ in accuracy for $N_{it} \simeq 40$. An analogous behaviour comes out also for $J^\pi = 2^+$, shown in Fig. \ref{fig:genEIG_b2}(e) for SA and in Fig. \ref{fig:genEIG_b2}(f) for QA. The error on the lowest eigenvector reflects the trend described for the eigenvalue error and it remains above $10^{-7}$ in all the cases.

Our results demonstrate a significant performance advantage of QA algorithm over SA also when solving GEVPs with $b=4$, as shown in Fig. \ref{fig:genEIG_b4}.\\
Here, QA achieves an eigenvalue accuracy of $10^{-8}$ within only $N_{it} \simeq 20$, while SA requires $N_{it} \simeq 90$ (Figures \ref{fig:genEIG_b4}(a) and (b) for $J^\pi = 0^+$). Similarly, QA finds the eigenvector with an accuracy of $10^{-7}$ for $N_{it} \simeq 60$ compared to $N_{it} \simeq 230$ needed by SA to reach the same level of accuracy.

Similar observations hold for $J^\pi = 1^-$ and $J^\pi = 2^+$ (Figures \ref{fig:genEIG_b4}(c-f)). While SA struggles to achieve 8-digit accuracy in these cases, requiring from $N_{it} \simeq 350$ to $N_{it} \simeq 1200$, QA readily converges to an accuracy of $10^{-8}$ within $N_{it} \simeq 20$ to $N_{it} \simeq 30$.

From the above analysis, we can conclude that in all examined cases, the use of QA over SA guarantees a quicker convergence to the $10^{-8}$ accuracy, since it requires less iterations. This gap in performance becomes more noticeable as the matrix size increases. In particular, there are cases in which SA doesn't even let reach the eigenvalue accuracy of $10^{-8}$, as shown in Figures \ref{fig:genEIG_b4}(c),(e) for $J^\pi = 1^-, 2^+$. In all the other cases, the boost in performances found ranges between $78\%$ and $98\%$ for the number of iterations, indicating  a clear improvement in sampling efficiency of QA with respect to SA.

\subsection{Full spectrum}\label{sec:FullSpectrum}
We now analyse the results for the full spectrum. Due to the limited Quantum Processing Unit (QPU) time available on the D-Wave quantum annealer, in this section we only demonstrate the reliability of our approach for the SA case. Here, the QUBO algorithm used to determine the dominant eigenpair (Algorithm \ref{max_magnitude_algorithm}) is combined with the three different deflation methods outlined in Section~\ref{sec:Matrix_Deflation_Techniques}.

\begin{figure}[h!]
    \centering
    \includegraphics[scale=0.47]{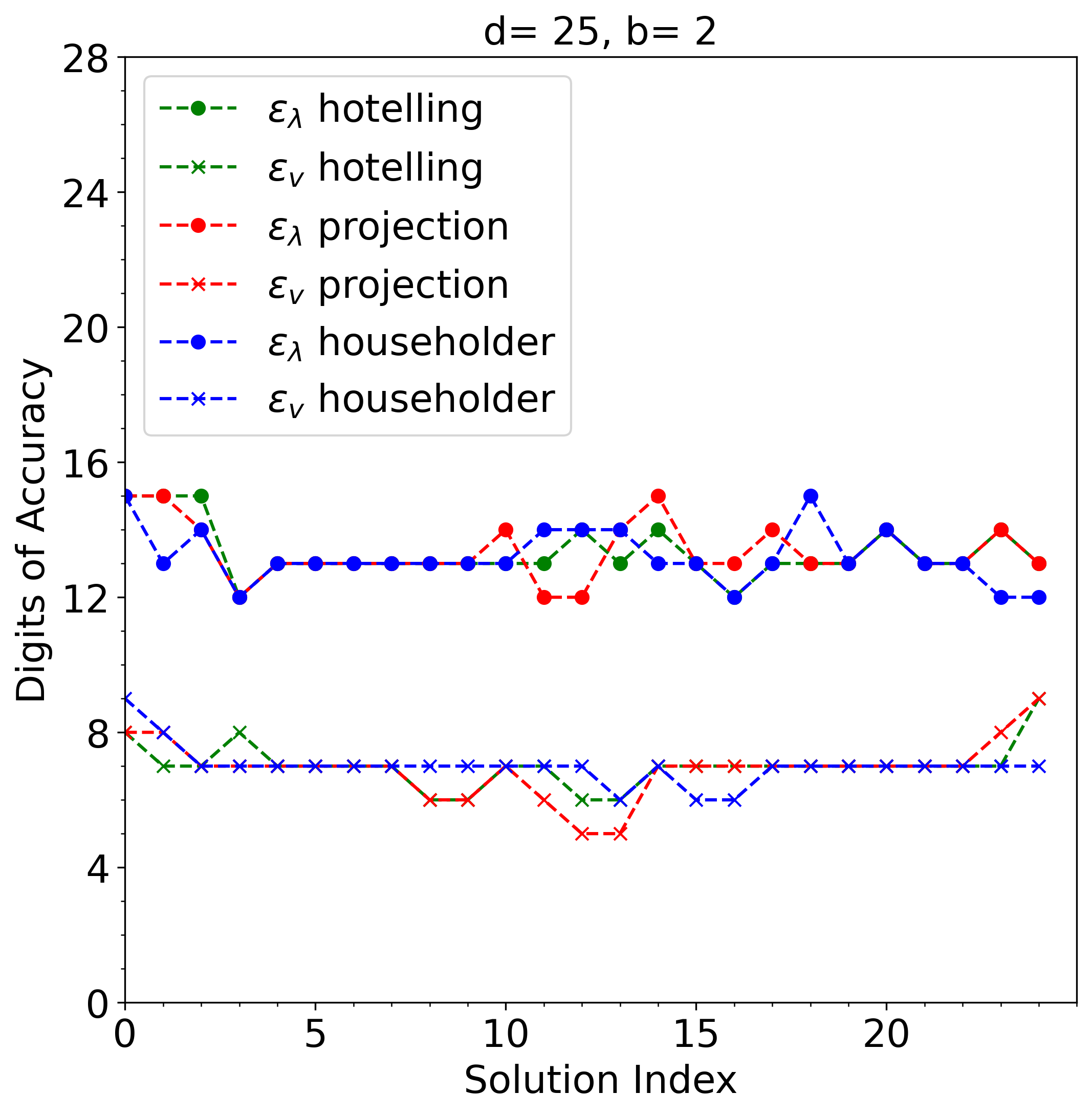}
    \caption{Digits of accuracy of the eigenpairs of the $J^\pi=0^+$ SEVP EMPM full Hamiltonian with respect to the three different methods adopted, with $b=2$. }
    \label{fig:SEVP_b2}
\end{figure}

\begin{figure}[h!]
    \centering
    \includegraphics[scale=0.47]{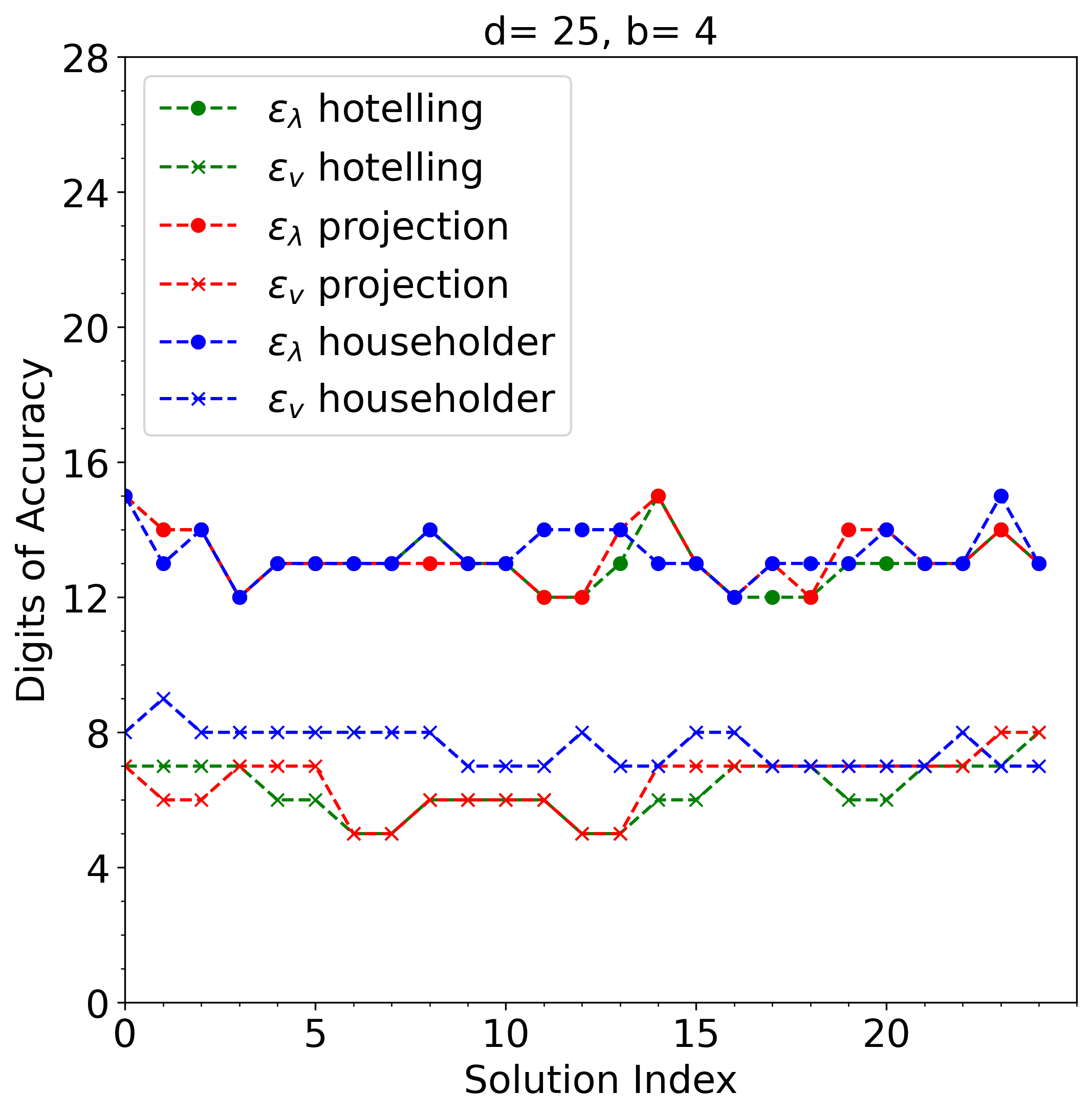}
    \caption{Same as Fig. \ref{fig:SEVP_b2}, for $b=4$.}
    \label{fig:SEVP_b4}
\end{figure}
We start by analysing the results obtained for the SEVP. Here, we report the results for the $J^\pi = 0^+$ EMPM full Hamiltonian with $b = 2$ (Fig.~\ref{fig:SEVP_b2}) and $b = 4$ (Fig.~\ref{fig:SEVP_b4}). A detailed inspection of the results shows that all the implemented deflation techniques yield highly accurate final eigenvalues and eigenvectors, achieving up to $13$ and $8$ correct digits, respectively. 

The total execution time ($te_t$), comprising both the deflation step and the annealing phase time, varies across the three deflation methods. For the Hotelling method, $te_t = 167.13\,$s with $b=2$ and it increases to $526.06\,$s with $b=4$. For the Orthogonal Projection method, $te_t = 155.47\,$s with $b=2$ and it increases to $533.20\,$s with $b=4$. Lastly, the Householder method is significantly faster, with $te_t = 85.53\,$s with $b=2$ and $te_t = 211.39\,$s with $b=4$. This performance advantage stems from its deflation strategy, which progressively reduces the size of the active submatrix, thereby minimizing computational effort as the algorithm advances.

\begin{figure}[h!]
    \centering
    \includegraphics[scale=0.47]{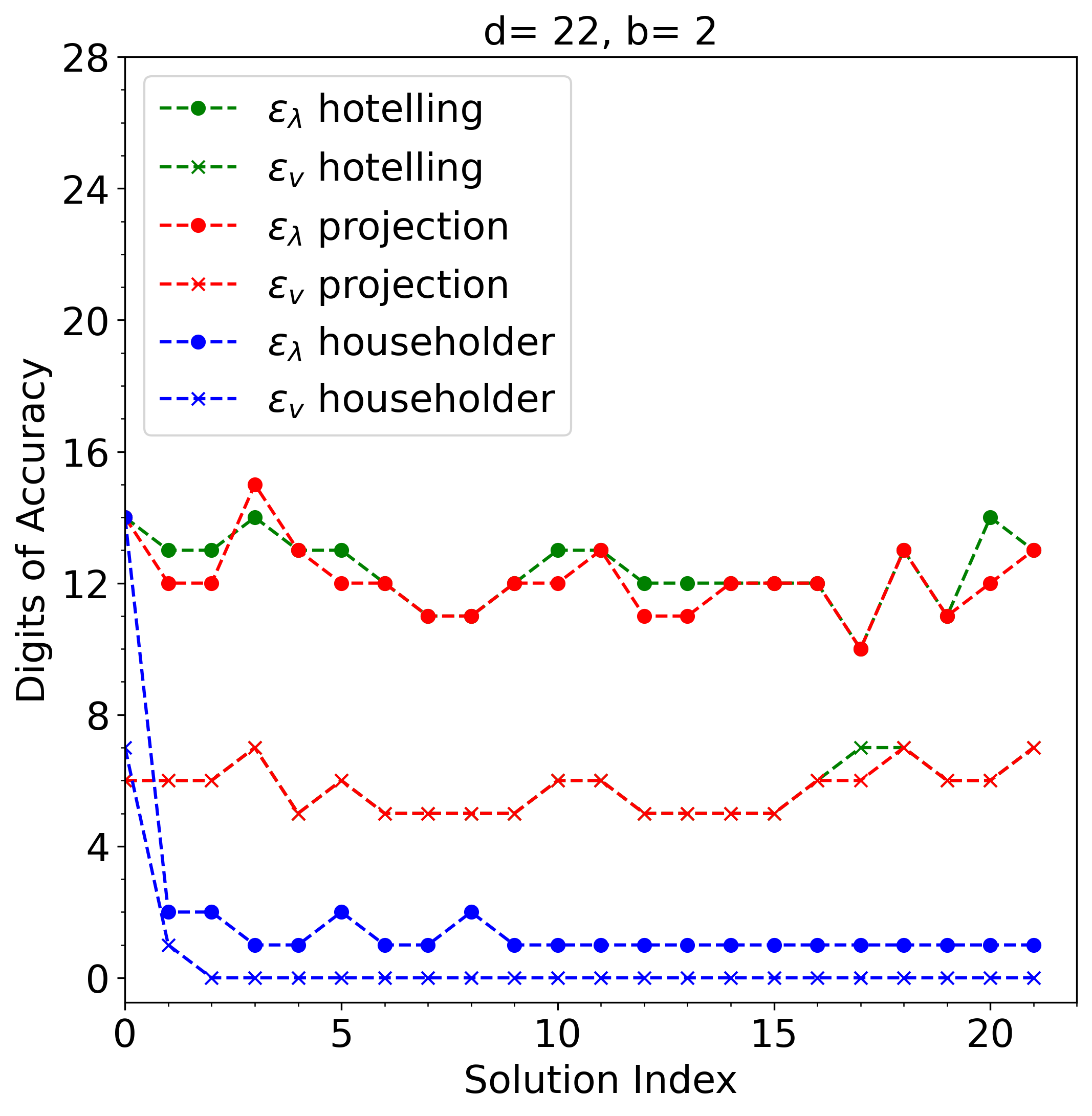}
    \caption{Digits of accuracy of the eigenpairs of the GEVP EMPM with 
$J^\pi = 0^+,1^-, 2^+$ with respect to the three different methods adopted, with $b=2$. }
    \label{fig:GEVP_b2}
\end{figure}

\begin{figure}[h!]
    \centering
    \includegraphics[scale=0.47]{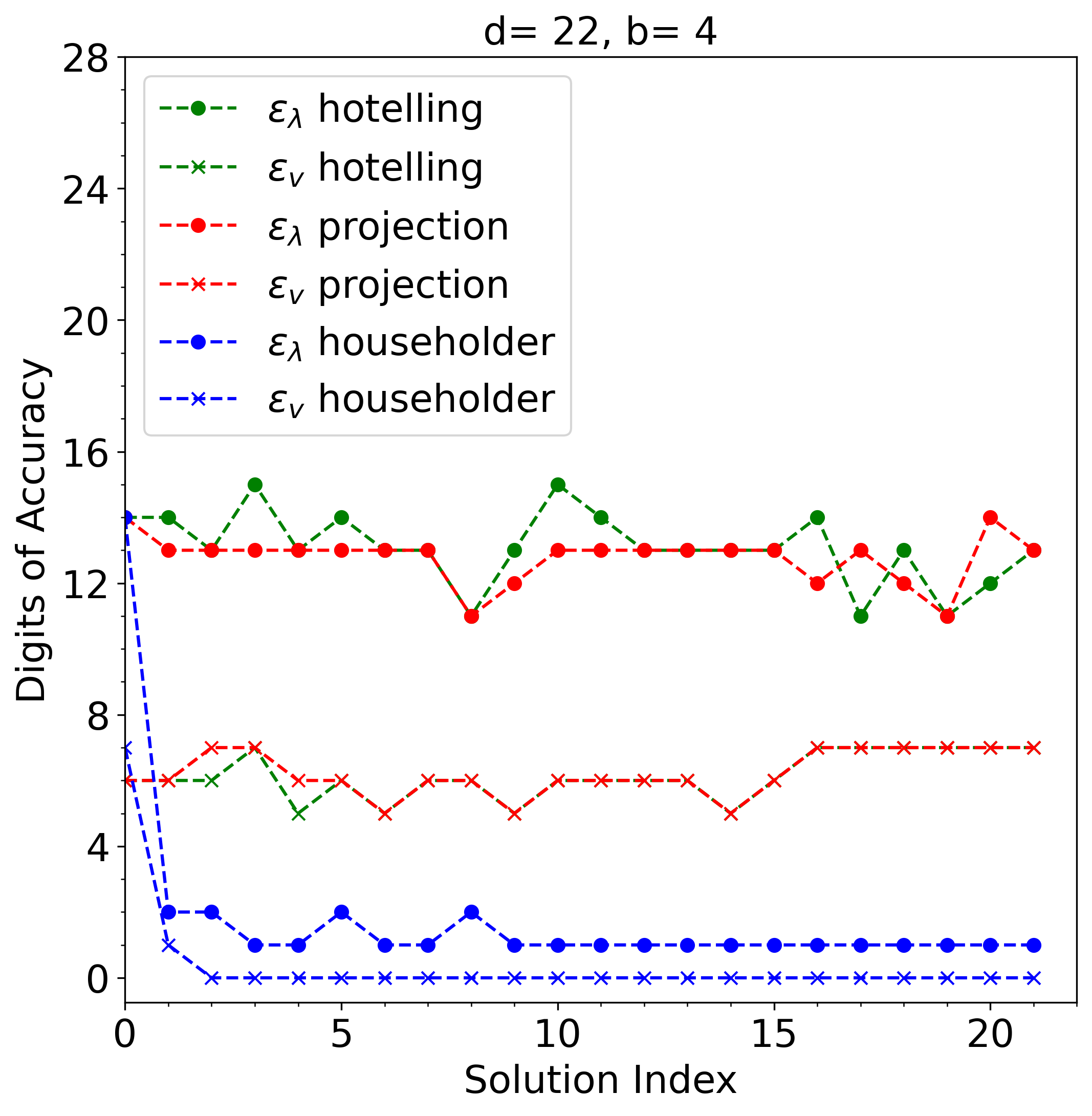}
    \caption{Same as Fig. \ref{fig:GEVP_b2}, for $b=4$.}
    \label{fig:GEVP_b4}
\end{figure}
In the case of the GEVP, at variance with the SEVP, the repeated application of Householder reflectors for deflation introduces a progressive deterioration of the matrix \(B\) that compromises numerical stability. Although each reflector
\begin{equation}
  H = I - 2 \frac{\ket{u}\bra{u}}{\braket{u\mid u}},
\end{equation}
is orthogonal, the transformations, reported in Eq. \eqref{HouseholderTransofrmations}, that it generates, induce fill-in and distort the original symmetry (see, e.g., \cite{Stewart2001,Mehl2003}). Furthermore, rounding errors introduced in the early deflations are amplified in the subsequent ones, preventing already deflated vectors from remaining normalized with respect to the \(B\)-inner product. Additionally, when eigenvalues lie close together \cite{Watkins2007}, the procedure tends to fail to converge to the target eigenvalue, since adjacent spectral values accentuate small perturbations in the basis and make it difficult to correctly isolate the desired eigenvalue. As a result, the Householder deflation procedure is unstable and not applicable in this phase of the numerical comparison.

The other two methods, Hotelling and Orthogonal Projection, instead, produce results with high accuracy, comparable to the standard case.

The total execution time ($te_t$) for the three deflation methods varies as follows. For the Hotelling method, $te_t = 963.11\,$s with $b=2$ and it increases to $1293.64\,$s with $b=4$. For the Orthogonal Projection method, $te_t = 914.56\,$s with $b=2$ and it increases to $1024.78\,$s with $b=4$. In contrast, the Householder method remains significantly faster, with $te_t = 181.22\,$s with $b=2$ and $te_t = 443.97\,$s with $b=4$.

\section{Conclusion}

In this manuscript we have applied a QUBO-based hybrid quantum-classical framework for computing the eigenspectrum of realistic EMPM Hamiltonians. The first part of the analysis focused on the ground-state estimation, where direct comparison between simulated annealing and quantum annealing reveals that the quantum approach yields systematically better results. This clearly highlights an advantage of quantum sampling with respect to classical simulated annealing.

Building upon the validated ground-state methodology, we have then extended the approach to full-spectrum reconstruction via an iterative deflation strategy. Our findings demonstrate that the proposed QUBO-based iterative scheme, combined with classical post-processing, accurately recovers all eigenpairs of the EMPM Hamiltonian.

In the SEVP case, all three deflation strategies-Hotelling, Orthogonal Projection and Householder-yield eigenvalues and eigenvectors with machine precision for both $b=2$ and $b=4$, confirming the numerical robustness of the method for the solution of symmetric eigenvalue problems. In contrast, for the GEVP, the progressive loss of orthogonality induced by repeated Householder updates leads to rapid numerical deterioration. Consequently, Householder deflation cannot be reliably employed in this context. The Hotelling and Orthogonal Projection methods, instead, preserve accuracy across all iterations and remain effective also in the generalized case.

Overall, these findings confirm that the proposed hybrid QUBO-deflation framework provides a viable route to extend quantum annealing methods beyond ground-state estimation toward systematic reconstruction of the entire spectrum. The comparison between simulated annealing and quantum annealing indicates a performance advantage of the quantum approach in exploring the energy landscape associated with the QUBO formulation of the EMPM Hamiltonian.
Although this result is obtained within the present QUBO framework and for matrix sizes compatible with current quantum annealing hardware, it supports the potential of quantum annealing as an effective strategy for spectral estimation problems of this class.

The demonstrated robustness of the deflation mechanisms establishes the required groundwork for future implementations fully executed on quantum annealers. In particular, a natural next step will be to reduce the degree of classical part of the algorithm by encoding also the deflation step directly as a QUBO minimization problem, with the long-term goal of realizing a fully quantum annealing-based spectral solver with minimal hybrid overhead.
\section*{Acknowledgments}
G.D.G. acknowledges the support from the EU-FESR,
PON Ricerca e Innovazione, Grant No. 2014-2020- DM
1062/2021. This work is also supported  the Czech Science Foundation (Czech Republic), P203-26-21972S. Computational
resources were partially provided by the e-INFRA CZ project
(ID:90254), supported by the Ministry of Education,
Youth and Sports of the Czech Republic, and by the
ELIXIR-CZ project (ID:90255).
\bibliography{EMPM_QA}
\end{document}